# Towards nonlinear electrodynamics without renormalization

Ho-Dong Jo[1] and Chol-Song Kim[1, 2]

[1]Department of Physics, Kim Il Sung University, Pyongyang 999093, D.P.R. Korea
[2]School of Physics, Peking University, Beijing 100871, P.R. China

e-mail: cskim@pku.edu.cn, 2102020201@stu.pku.edu.cn

**Abstract**

In this paper is considered nonlinear electrodynamics (NE) which does not satisfy the linear superposition principle (LSP). Since the presentation of the special theory of relativity, it has been commonly accepted that a famous formula $E = mc^2 = m_0 c^2/\sqrt{1 - v^2/c^2}$ expresses the energy of a free particle only. However, while studying the experiment for the annihilation of particle and antiparticle and the production of the photon in terms of the law of energy conservation of particle and field, we obtain a conclusion that $E$ includes not only the energy of a free particle but also the energy of its self –fields (electromagnetic field and gravitational field), that is, all fields produced by the particle: for detail, see Sect. 2, here the argument is confined to the electromagnetic field only. Hence, a formula for the energy in the special theory of relativity comes to have a more inclusive meaning than Einstein had thought of it. However, unfortunately, on computing the total energy of a free particle and its self-field (electrostatic field), the Maxwell theory gives the infinite divergence of the energy of an electrostatic field, from which radiation reaction also leads to the divergence of the same form. In this context, the renormalization procedure has been introduced to remove the divergence, in which the mass entered the formula for energy in the special theory of relativity is regarded as 'bare mass' devoid of field. But, from the above-mentioned conclusion, it follows that the measured mass $m(m;$ motion mass)in the formula intrinsically includes the mass of field from the outset and so the renormalization procedure results in loss of physical meaning. This implies another important conclusion, i.e., the Maxwell theory should be modified such that the total energy of a free particle and its fields naturally equals $mc^2$ without invoking the renormalization procedure. Based upon such an idea and the correspondence principle, through introducing a new non-Euclidean four-dimensional space, so-called KR space in which the metric is the function of coordinates and 4-velocity, we reformulate the action function to be a function defined in the nonlinear version and from it find equation system for a charge and the field with a modified version, which is reduced to that in the Maxwell theory under the weak field condition: here, weak field condition is given by $|U/m_0 c^2| \ll 1$, $U$; interactional energy. With the equation system, we show that the divergence problem arising from the Maxwell theory, without invoking renormalization, is very naturally solved by sensible mathematics and then demonstrate that the total energy of a free charge and its field equals $mc^2$ and so the energy of an electrostatic field leads to the finite. In this course are derived the modified version of Coulomb potential and of radiation reaction without renormalization which contains an interference effect (a nonlinear effect that does not obey LSP). In addition, a modified version for the energy of a particle gives the relation between inertial mass and energy in a general form, i.e., so-called "the equivalence of inertial mass and energy", which depends on the interaction of a particle with the external field and in the absence of the external field is reduced to the relation between inertial mass and energy in the special theory of relativity. More interesting is that the formula for the energy specifies a mechanism of the annihilation and production of new particles based on the conservation law of energy, while the formula or the relation of mass and energy in the special theory of relativity indicates the existence of a new particle, i.e., photon: the pair annihilation of electron and positron and the production of photons, the capture of an electron by a neutron. The theoretical result shows good agreement with experiments under the strong field condition ($|U/m_0 c^2| \cong 1$, $U$; interaction energy), which cannot be explained clearly in terms of the Maxwell theory. Based on the classical non-linear electrodynamics without renormalization, we briefly touch on the quantum nonlinear electrodynamics in which a modified version of the Dirac equation is derived and divergence rising from the S-matrix of quantum electrodynamics is canceled without renormalization. The modified version of Coulomb potential predicts a new correction in the scattering of electrons by a nucleus, which should be verified by experiment in the future.

## 1. Introduction

Finding the equation of motion for a charged particle in the classical radiation field is a very old problem [1-4]. When trying to consistently derive the effective equation of motion for a point charge, one comes up against the inevitable infinities or divergences arising due to taking a charged particle to be a point: an infinity of electrostatic field energy (self-field energy) and divergent contribution in radiation reaction. From this, the question arises whether the Maxwell theory is a consistent one in physics [5]. In order to solve this problem, different remedies have been explored [6]. The first approach is to modify the Maxwell theory as has been done by Born and Infeld [7] or Podolsky and Schwed [8]. They recently regained attention [9]. The second approach is to introduce an extended charge model as has been done by Abraham [10], Lorentz [11], and many others [12]; see [13] for a recent report. The third approach is the renormalization procedure which was first proposed by Dirac [14] and has been studied in different forms; see [15-17]. If we confine our consideration to take a charge to be a point, one of them, the first and third approaches, should be adopted. But it is not so satisfactory to conclude that the renormalization is a reasonable approach. The renormalization procedure violates the scientific requirement that there is a unique description for any given physical fact: subtracting the infinite (mass of self-field) from another infinite (bare mass) is not sensible mathematics and gives no unique mathematical value [18,19] (Dirac rejected strongly the renormalization procedure in the quantum electrodynamics though he was a founder of renormalization program in the classical electrodynamics). The version of function of bare mass is not defined unambiguously and that subtracting mass of self-field from bare mass leads to an unambiguous mass (renormalized mass) measured is only assumed but not testified. The additional comment is that, in Maxwell's electrodynamics, by means of the renormalization procedure, the essential meaning of the basic concept (mass of a particle) defined in the first of the building of a theory is, in the middle of the evolution of the theory, basically modified and redefined. This shows that the renormalization procedure stands against a scientific rule to be naturally introduced for the building of a consistent theory according to which basic concepts of the theory should always preserve their own meaning in the whole area of a theory. This divergence problem in classical electrodynamics reappears in a more serious form at the stage of quantum electrodynamics (QED) [20]. In this context, there were some attempts to build QED without renormalization [21], but by a preponderance of the renormalization procedure, these were suppressed. In this paper, we consider the classical nonlinear electrodynamics (CNE) without renormalization to construct classical electrodynamics consistently and then, based upon it, touch the quantum nonlinear electrodynamics (QNE) without renormalization. For this purpose, we present a starting idea or starting postulate that the total energy of a free particle and its self-field equals $mc^2$: it will be discussed in Sect.2. Until now, the general consensus would likely be that the Maxwell theory is a complete one, and for the elimination of the divergence is no alternative to introduce the renormalization procedure within a framework of the Maxwell theory as far as a particle is regarded as a point. Therefore, finding nonlinear electrodynamics (NE) without renormalization is beyond the hitherto thought in physics. But, our starting idea contradicts strongly renormalization and presents the limitation of the Maxwell theory. From such an unsatisfactory situation, we are motivated to write this paper. The importance of the establishment of NE is not purely confined only to the elimination of the divergence, while NE necessarily gives rise to nonlinear effects which cannot be described in terms of the Maxwell theory. It is anticipated that the study for it would be contributed to a new advance in electrodynamics.

The plan of the paper is as follows. Sect. 2 puts forward starting postulate I and in Sect. 3 is formulated the action function for the motion of a charge and discussed the equations for the motion of a charge. Sect. 4 studies the experimental verification for the theoretical results under the strong field condition, obtained from the equations. In Sect. 5 is considered the generalized conservation law of charge and application limit of the gauge symmetry. Sect. 6 studies the energy-momentum tensor and field equations. Sect. 7 introduces KR space (new non-Euclidean space) on which Lagrangian is formulated, and gives physical analysis and normalization rules for main functions defined on KR space. Sect. 8 discusses the main results of CNE: modified versions of Coulomb's law and radiation reaction, and total energy of a charge and its self-field which is considered without renormalization. In Sect. 9, we briefly touch quantum nonlinear electrodynamics (QNE) in which a modified version of the Dirac equation is derived and the divergence problem coming from the S-matrix of quantum electrodynamics is naturally solved without renormalization. Then we present that consideration for the scattering of an electron by the modified version of Coulomb potential produced by a nucleus yields a new correction to be verified by experiment in the future.

## 2. Starting postulate I: basic idea for the total energy of a particle and its self-field

Historically, finding classical nonlinear electrodynamics was motivated by the credit for the 'principle of finiteness' by which a satisfactory theory should be the absence of divergences in physical quantities [7-9]. Unfortunately, even to the present day, there is not yet a leading principle for building a unique theory, a basic principle that enables one to satisfy the 'principle of finiteness' and specifies intrinsic criteria of validation or reputation for attempts at the theory.

In this section, we present a starting postulate I (providing a candidate of the leading principle) that underlies CNE, which can allow one to define the total energy of a charged point particle and its self-field as the finite: in this case a particle is taken to be a point. For this purpose, we consider the annihilation of particle-antiparticle well known in physics. The formula of energy conservation can be written as follows:

$$\frac{m_0 c^2}{\sqrt{1 - v_1^2/c^2}} + \frac{m_0 c^2}{\sqrt{1 - v_2^2/c^2}} = 2\hbar\omega, \qquad (1)$$

where $m_0 c^2/\sqrt{1 - v_1^2/c^2}$ is the energy of a free particle and $m_0 c^2/\sqrt{1 - v_2^2/c^2}$ the energy of a free antiparticle, which is the main result of the special theory of relativity (SR): $v_1$ is the velocity of the particle and $v_2$ the velocity of the antiparticle. Since Einstein presented the special theory of relativity(SR), $mc^2$ has been considered only as the energy confined to the particle. When describing Lagrangian for the motion of a particle in the electromagnetic field, the rest mass $m_0$ that enters the particle-related part was regarded as "bare mass" devoid of field. On the other hand, while considering radiation reaction, in order to eliminate the divergence of self-field energy, "dressed mass" was redefined by the renormalization procedure [14]. But, when considering Eq. (1), from the outset, it should have been viewed that $E = mc^2 = m_0 c^2/\sqrt{1 - v^2/c^2}$ includes not only the energy of a particle but also the energy of the field (self-field energy) produced by the particle. Why should it be viewed in that way? It, in a word, is based upon the fact according to which the total energy of particles and fields created by them should be always conserved. In the case of considering $mc^2$ to be energy confined to a particle only, the conservation formula of the total energy of particles and fields can be represented as

$$\left(\frac{m_0 c^2}{\sqrt{1-v_1^2/c^2}} + \varepsilon_1\right) + \left(\frac{m_0 c^2}{\sqrt{1-v_2^2/c^2}} + \varepsilon_2\right) = 2\hbar\omega + \varepsilon, \qquad (2)$$

where $\varepsilon_1$ is the energy of electromagnetic field and gravitational field created by the particle, and $\varepsilon_2$ the energy of two fields created by the antiparticle: in a general sense, we assume that particles produce the electromagnetic field, as well as the gravitational field. From (1) and (2), we get

$$\varepsilon_1 + \varepsilon_2 = \varepsilon, \qquad (3)$$

where $\varepsilon$ is the energy of another matter which appears newly except photon after the annihilation of a system of particle-antiparticle. However, until now has not yet been found any experimental data which, except photon, another matter occurred. Therefore, if there is something except photons after the annihilation of particle-antiparticle, there is no alternative to conclude that only the fields remain as it is, as an unmeasured invariant. On the other hand, as the particle-antiparticle is annihilated, the mass $m_0$ and charge $e$ also vanish, and thus the static gravitational field and electrostatic field created by the rest mass and charge also should disappear. Therefore, the result is

$$\varepsilon_1 + \varepsilon_2 \neq \varepsilon, \qquad (4)$$

with

$$\begin{cases} \varepsilon = 0 \\ \varepsilon_1 + \varepsilon_2 > 0 \end{cases} \qquad (5)$$

Consequently, when considering $mc^2$ only as the energy confined to a particle, with the occurrence of photons, we lead to the conclusion that the energy of electromagnetic field and gravitational field should vanish, which stands against the conservation law of energy. In this context, we define starting postulate I as follows: *for a system consisting of a free particle and its field, $mc^2$ ($m$ is the measured mass; $m = m_0$ in the rest state, $m = m_0/\sqrt{1-\beta^2}$ in the motion state) is the same as the total energy of a particle and all fields produced by it (electromagnetic field plus gravitational field) but not the energy of the particle only:* here $\beta = v/c$ ($v$; velocity of a particle) and $m$ is the total mass of a particle and its self-field. Throughout the paper, $\beta$ denotes $v/c$.

In this paper, the fact that the total energy of a free particle and fields created by it is just equal to $mc^2$ is regarded as a new starting postulate or starting point for developing CNE. This shows that the formula for the energy of a free particle in SR involves more inclusive and universal physical meaning than Einstein had thought of it: an inclusive concept of the total energy of a system that consists of a particle and its self-fields. From the starting postulate I, we can obtain the following conclusions. The starting postulate I implies that in the paper, $E$ (the total energy of a free particle and its field) obtained from energy-momentum tensor should lead to

$$\begin{cases} E = m_0 c^2/\sqrt{1-\beta^2} \; + \; U_0 = mc^2 \\ U_0 = 0 \end{cases} \qquad (6)$$

where $U_0$ is the energy of the self-field of a particle, which gives no divergence, being zero according to the starting postulate I: it is because $mc^2$ contains not only the energy of a particle but also the energy of its self-field. In this case, it should be emphasized that the system consisting of a particle and its field should be considered as integrity and so a question about what share of $mc^2$ is divided into the energy of matter and self-field, respectively and whether the bare mass and the mass of self-field are finite or infinite are, in principle, invalid. Therefore, the paper does not need the definitions of bare mass and mass of self-field, and thus, the introduction of the renormalization procedure.

When one considers a system of many particles, the total energy of $n$ particles and their fields (self-field plus interactional field), $E_n$, should arrive at

$$\begin{cases} E_n = \sum_{i=1}^{n} m_i c^2 + \sum_{i=1}^{n} U_i \\ U_i = \sum_{j=1(i\neq j)}^{n-1} U_{ij}\left(\sim e_i e_j / R_{ij}\right), \qquad U_{ii} = 0 \end{cases} \tag{7}$$

where $m_i c^2$ is the energy of the $i$th particle and its self-field, $U_i$ the interaction energy of the $i$th particle with the whole fields produced by other particles, $U_{ij}$ the interaction energy of different two particles, $R_{ij}$ the distance between arbitrary two particles, and the self-field energy $U_{ii}$ should vanish. Hence, CNE should be built such that the infinite terms arising from self-field vanish naturally and only the terms of interactional energy dependent on the mutual distribution of particles remain. This explicitly shows that starting postulate I is a leading principle that enable one to satisfy the 'principle of finiteness'. However, unfortunately from the Maxwell theory, we cannot find the conclusion obtained above. If one tries to get such a conclusion, it is no alternative to introduce the renormalization procedure. But as shown above, the starting postulate I does not need the renormalization procedure and stands strongly against it. This implies a drawback of the Maxwell theory. In the view of the correspondence principle, a new theory should include the former theory as an approximate version. the Maxwell theory is a linear theory that obeys LSP. In a general sense, a nonlinear theory comprises a linear theory as an approximate version as if the non-Euclidian geometry includes the Euclidian geometry, that is, linear geometry as an approximate version (within an infinitesimal area of space) Therefore, from starting postulate I and correspondence principle, we can conclude that true classical electrodynamics must be a nonlinear theory that includes the Maxwell theory as an approximate version: staring postulate II is discussed in Subsect.7.2.

## 3. Equation of motion of a charge in CNE

In this section, we formulate the action for the motion of a charge and derive the equation of motion of a charge in the electromagnetic field. In CNE, the action for the motion of a particle should satisfy the following requirement. First, the action function should be invariant with respect to the Lorentz transformation from Einstein's postulates in SR. Second, the action must constitute a nonlinear function which, according to the principle of correspondence, includes the linear Lagrangian in the Maxwell theory as an approximate version. Third, $mc^2$ in the action function for a free particle should be the total energy of a particle and its self-field: this is subject to the starting postulate I argued in Sect. 2 and so from the actions for a free particle and field produced by it should follow that the total energy of a particle and its self-field is the same as $mc^2$, which is derived from the energy-momentum tensor obtained from the action.

From the above requirement we define the action for the motion of a charge as follows:

$$S = \int L ds = -m_0 c \int ds \ (g_{\mu\nu} u^\mu u^\nu)^{\frac{1}{2}} = -m_0 c \int [\delta_{\mu\nu}(1 + 2\alpha A_\lambda u^\lambda) u^\mu u^\nu]^{\frac{1}{2}} ds, \qquad (8)$$

with $L = -m_0 c (g_{\mu\nu} u^\mu u^\nu)^{1/2}$ and $g_{\mu\nu} = \delta_{\mu\nu}(1 + 2\alpha A_\lambda u^\lambda)$, where $\delta_{\mu\nu}$ is the Minkowski metric tensor given by $\delta = \text{diag}(1, -1, -1, -1)$, $A_\lambda$ the 4-dimensional electromagnetic field vector, $u^\lambda$ the 4-dimensional velocity, $A_\lambda u^\lambda = \delta_{\lambda\sigma} A^\lambda u^\sigma$, $\alpha$ a constant defined from the approximate condition $\alpha = e/m_0 c^2$, $A_\lambda = A_{ex} + A_{in}$ ($A_{ex}$; external field and $A_{in}$; self-field of a particle ), and $g_{\mu\nu}$ is a non-Euclidean metric tensor (a metric tensor of KR space dependent on the velocity 4-vector and coordinates of space-time: see 7.1 for KR space and as will be shown in Sect. 8, $A_{in} = 0$ holds for a free particle and $A_{in}$ is the small quantity of order $e^2/m_0 c^3$ in the presence of radiation reaction. Hence, from the outset in the action (8), the interactions of a particle with an external field as well as self-field are treated in a unified framework, while the Lagrangian in the Maxwell theory includes only the external field due to the divergence of the self-field potential and then, with the aid of a very artificial approach of renormalization, the interaction of a particle with its self-field is, additionally from the outside of a logical system of the theory, introduced into the equation of motion, which finally leads to the Lorentz-Abraham-Dirac equation (LAD equation) [6]. As will be seen in Sect. 8, describing $A_\lambda$ in a unified framework of $A_{ex}$ and $A_{in}$ is of paramount importance for a consistent solution to the divergence problem.

The action function (8) is conformed to the requirement mentioned above. First, (8) is invariant with respect to Lorentz transformation: as can be easily demonstrated, because in the metric tensor $g_{\mu\nu}$, $\delta_{\mu\nu}$ is the Minkowski metric tensor and $A_\lambda u^\lambda$ the scalar product of the field 4-vector and the velocity 4-vector, (8) is invariant with respect to Lorentz transformation. Second, as $\alpha A_\lambda u^\lambda$ is in the square root, the term characterizing interaction becomes a nonlinear version and then gives birth to the interference effect of interaction, and so, the linear superposition principle (LSP), as well as gauge symmetry is broken. The gauge symmetry is a basic principle that underlies present classical and quantum electrodynamics. In this relation, serious problems may arise why gauge symmetry necessarily should be broken and if so, what becomes of the conservation law of charge, what is the associated conservation law: this is intensively argued in Sect. 5. Unfortunately, with one or several phrases can never be given a persuadable description of the problem. Throughout the paper, we have confirmed that gauge symmetry and LSP are the fundamental factors giving rise to divergence problems in classical and quantum linear electrodynamics, and only classical and quantum nonlinear electrodynamics described by logical and sensible mathematics ensures the finiteness of physical quantities. It is obvious that while considering CNE with starting postulate I and correspondence principle, the breaking of gauge symmetry and LSP are inevitable because the two principles come from the linearity of interaction Lagrangian. Now, allowing for the approximation condition $r \gg r_0$ ($r$; the interaction radius, $r_0$; around electron radius, $\sim 10^{-13}$ cm), the action (8) is transformed into the action for the motion of a charge in Maxwell's electrodynamics. The action (8) can be rewritten as

$$S = -m_0 c^2 \int dt \left(1 - \frac{V^2}{c^2}\right)^{\frac{1}{2}} (1 + 2\alpha A_\lambda u^\lambda)^{\frac{1}{2}}. \qquad (9)$$

Allowing $|2\alpha A_\lambda u^\lambda| \ll 1$, a Taylor series expansion of the action (9) arrives at

$$S \approx -m_0 c^2 \int dt \left(1 - \frac{V^2}{c^2}\right)^{\frac{1}{2}} - m_0 c^2 \alpha \int A_\lambda \dot{x}^\lambda \, dt + \frac{1}{2} m_0 c^2 \alpha^2 \int dt \left(A_\lambda \dot{x}^\lambda\right)^2 \left(1 - \frac{V^2}{c^2}\right)^{\frac{1}{2}}. \tag{10}$$

Neglecting the third term with the coefficient $\alpha^2$ ($1/c^4$) in (10) follows

$$S \approx -m_0 c^2 \int dt \left(1 - \frac{V^2}{c^2}\right)^{\frac{1}{2}} - m_0 c \alpha \int A_\lambda \dot{x}^\lambda \, dt. \tag{11}$$

Substituting $\alpha = e/m_0 c^2$ in (11), we have

$$S \approx -m_0 c^2 \int dt \left(1 - \frac{V^2}{c^2}\right)^{\frac{1}{2}} - \frac{e}{c} \int A_\lambda dx^\lambda. \tag{12}$$

The action (12) coincides with the action in the Maxwell theory. If so, what is the physical meaning implied by the approximate condition $|2\alpha A_\lambda u^\lambda / m_0 c^2| \ll 1$? With $2\alpha A_\lambda u^\lambda \approx 2\alpha \varphi u^0 \approx 2\alpha \varphi$ ($\varphi$; Coulomb potential) in the approximation ignoring high order terms more than $1/c^2$, $|2\alpha A_\lambda u^\lambda / m_0 c^2| \ll 1$ leads to $|2e\varphi / m_0 c^2| \ll 1$ (the condition of weak field). When considering the motion of an electron under the electrostatic field produced by a proton, we have

$$2\alpha\varphi = 2\frac{e}{m_0 c^2} \cdot \frac{e}{r} = \frac{2e^2}{m_0 c^2} \cdot \frac{1}{r} = \frac{r_0}{r}, \tag{13}$$

where $r_0 = 2e^2/m_0 c^2$ is the electron radius ($\approx 10^{-13}$cm). Consequently, the approximate condition by which is transformed into the Lagrangian for the motion of a particle in Maxwell's electrodynamics corresponds to the case where $|2e\varphi / m_0 c^2| \ll 1$ (the condition of weak field) or $r \gg r_0$ ($r$; the interaction distance between particles) holds. In fact, all experiments put in the ground of the Maxwell theory were conducted under some approximate condition, i.e. a macroscopic region satisfying $|2\alpha A_\lambda u^\lambda / m_0 c^2| \approx |2e\varphi / m_0 c^2| \ll 1$ and then, in the process of systemizing and generalizing experimental results by an inductive method, Maxwell's electrodynamics was built. However, without some special evidence, one cannot conclude that experiments would be valid also under the strong field condition $|2\alpha A_\lambda u^\lambda / m_0 c^2| \approx |2e\varphi / m_0 c^2| \approx 1$ or $r_0/r \approx 1$. The actions (8) and (9) are valid in not only a region far from $r_0$ but also the neighborhood of $r_0$ or even a region satisfying $r = r_0$: e.g., a region where the annihilation of electron and positron arises.

In the action (8), we can obtain the equation of motion of a charge from the variation principle

$$\delta S = \int ds \left[\frac{\partial L}{\partial x^\lambda} \cdot \delta x^\lambda + \frac{\partial L}{\partial u^\lambda} \cdot \delta\left(\frac{dx^\lambda}{ds}\right)\right] = 0. \tag{14}$$

When computing variation of 4-velocity in (14), it should be taken into account that $ds$ is a function of 4-velocity: in KR space, 4-velocity is the implicit function. Thus, the variation of 4-velocity is calculated as follows:

$$\frac{\partial L}{\partial u^\lambda} \cdot \delta\left(\frac{dx^\lambda}{ds}\right) = \frac{\partial L}{\partial u^\lambda}\left[\frac{d\delta x^\lambda}{ds} + \delta\left(\frac{1}{ds}\right)dx^\lambda\right]$$

$$= \frac{\partial L}{\partial u^\lambda}\left[\frac{d\delta x^\lambda}{ds} + \frac{\partial}{\partial u^\mu}\left(\frac{1}{ds}\right)\delta\left(\frac{dx^\mu}{ds}\right)dx^\lambda\right]$$

$$= \frac{\partial L}{\partial u^\lambda}\left[\frac{d\delta x^\lambda}{ds} - \alpha\gamma A_\mu u^\lambda \delta\left(\frac{dx^\mu}{ds}\right)\right]$$

$$= \frac{\partial L}{\partial u^\lambda}\left[\frac{d\delta x^\lambda}{ds} - \alpha\gamma A_\mu u^\lambda \frac{d\delta x^\mu}{ds} + (\alpha\gamma)^2 (A_\sigma u^\sigma) u^\lambda A_\mu \frac{d\delta x^\mu}{ds}\right.$$

$$\left. - (\alpha\gamma)^3 (A_\sigma u^\sigma)^2 u^\lambda A_\mu \left(\frac{d\delta x^\mu}{ds} + \delta\left(\frac{1}{ds}\right)dx^\mu\right)\right]$$

(15)

with

$$\delta\left(\frac{1}{ds}\right)dx^\lambda = \frac{\partial}{\partial u^\mu}\left(\frac{1}{ds}\right)\delta\left(\frac{dx^\mu}{ds}\right)dx^\lambda = -\alpha\gamma A_\mu u^\lambda \delta\left(\frac{dx^\mu}{ds}\right) = -\alpha\gamma A_\mu u^\lambda \left[\frac{d\delta x^\mu}{ds} + \delta\left(\frac{1}{ds}\right)dx^\mu\right]$$

$$= -\alpha\gamma A_\mu u^\lambda \frac{d\delta x^\mu}{ds} + (-\alpha\gamma A_\mu u^\mu)^2 A_l u^\lambda \left[\frac{d\delta x^l}{ds} + \delta\left(\frac{1}{ds}\right)dx^l\right]$$

where $\gamma = (1 + 2\alpha A_\lambda u^\lambda)^{-1}$ and $dS = \left(1 - \frac{v^2}{c^2}\right)^{\frac{1}{2}}(1 + 2\alpha A_\lambda u^\lambda)^{\frac{1}{2}}dt$. Performing continuously the above computation, we have

$$\frac{\partial L}{\partial u^\lambda} \cdot \delta\left(\frac{dx^\lambda}{ds}\right) =$$

$$= \frac{\partial L}{\partial u^\lambda}\frac{d\delta x^\lambda}{ds} - \alpha\gamma \frac{\partial L}{\partial u^\lambda}u^\lambda A_\mu \frac{d\delta x^\mu}{ds}[1 + (-\alpha\gamma A_\sigma u^\sigma) + (-\alpha\gamma A_\sigma u^\sigma)^2 + \cdots + (-\alpha\gamma A_\sigma u^\sigma)^n] +$$

$$+ (-\alpha\gamma)^{n+1}(A_\sigma u^\sigma)^n \frac{\partial L}{\partial u^\lambda}u^\lambda A_\mu dx^\mu \delta\left(\frac{1}{ds}\right)$$

(16)

Then, (16) arrives at

$$\frac{\partial L}{\partial u^\lambda} \cdot \delta\left(\frac{dx^\lambda}{ds}\right) =$$

$$= \lim_{n \to \infty}\left\{\left[\frac{\partial L}{\partial u^\lambda}\frac{d\delta x^\lambda}{ds} - \alpha\gamma \frac{\partial L}{\partial u^\lambda}u^\lambda A_\mu \frac{d\delta x^\mu}{ds}\sum_{n=0}^{\infty}(-\alpha\gamma A_\sigma u^\sigma)^n\right] + (-\alpha\gamma)^{n+1}(A_\sigma u^\sigma)^n \frac{\partial L}{\partial u^\lambda}u^\lambda A_\mu dx^\mu \delta\left(\frac{1}{ds}\right)\right\}$$

(17)

Now, assuming that $|-\alpha\gamma A_\sigma u^\sigma| < 1$, (17) results in

$$\frac{\partial L}{\partial u^\lambda} \cdot \delta\left(\frac{dx^\lambda}{ds}\right) = \frac{\partial L}{\partial u^\lambda}\frac{d\delta x^\lambda}{ds} - \frac{\alpha\gamma \frac{\partial L}{\partial u^\mu} u^\mu A_\lambda \frac{d\delta x^\lambda}{ds}}{1 + \alpha\gamma A_\sigma u^\sigma} \tag{18}$$

Substitution of (18) in (14) yields

$$\delta S = \int ds \left[\frac{\partial L}{\partial x^\lambda} \cdot \delta x^\lambda + \frac{\partial L}{\partial u^\lambda}\frac{d\delta x^\lambda}{ds} - \alpha\gamma \frac{\partial L}{\partial u^\mu}u^\mu \frac{A_\lambda}{1 + \alpha\gamma A_\sigma u^\sigma}\frac{d\delta x^\lambda}{ds}\right] = 0. \tag{19}$$

Integrating the second and third terms of (19) by parts and rearranging them, we have

$$\delta S = \int ds \delta x^\lambda \left[\frac{\partial L}{\partial x^\lambda} - \frac{d}{ds}\left(\frac{\partial L}{\partial u^\lambda} - \alpha\gamma \frac{\partial L}{\partial u^\mu}u^\mu \frac{A_\lambda}{1 + \alpha\gamma A_\sigma u^\sigma}\right)\right] = 0. \tag{20}$$

From (20) follows

$$\frac{\partial L}{\partial x^\lambda} - \frac{d}{ds}\left(\frac{\partial L}{\partial u^\lambda} - \alpha\gamma \frac{\partial L}{\partial u^\mu}u^\mu \frac{A_\lambda}{1 + \alpha\gamma A_\sigma u^\sigma}\right) = 0 \tag{21}$$

Eq. (21) gives the equation of motion of a charge: when obtaining (18), we have assumed $|-\alpha\gamma A_\sigma u^\sigma| < 1$. Eq. (21) holds true in the whole region of electromagnetic interaction including the strong field condition: $|-\alpha\gamma A_\sigma u^\sigma| \approx |\alpha A_\lambda u^\lambda|$ and under the strong field condition, since $|2\alpha A_\lambda u^\lambda| \approx 1$ holds, $|-\alpha\gamma A_\sigma u^\sigma| < |2\alpha A_\lambda u^\lambda|$ and so $|-\alpha\gamma A_\sigma u^\sigma| < 1$ is also valid.

In Eq. (21), 4-momentum is defined by

$$P_\lambda = -\left(\frac{\partial L}{\partial u^\lambda} - \alpha\gamma \frac{\partial L}{\partial u^\mu}u^\mu \frac{A_\lambda}{1 + \alpha\gamma A_\sigma u^\sigma}\right) \tag{22}$$

Calculating (22) yields

$$P_\lambda = m_0 c u_\lambda + \frac{e\gamma}{c}A_\lambda - \left(\frac{e\gamma\gamma'}{c} + \frac{e\gamma A_\sigma u^\sigma}{m_0 c^2}\frac{e\gamma\gamma'}{c}\right)A_\lambda =$$

$$= m_0 c u_\lambda + \frac{e\gamma}{c}A_\lambda - \frac{e\gamma\gamma'}{c}\left(1 + \frac{e\gamma A_\sigma u^\sigma}{m_0 c^2}\right)A_\lambda == m_0 c u_\lambda + \frac{e\gamma}{c}A_\lambda - \frac{e\gamma}{c}A_\lambda \tag{23}$$

where $\gamma' = (1 + \alpha\gamma A_\sigma u^\sigma)^{-1}$ and on computing (22) are employed the following expressions:

$$g_{\mu\nu}u^\mu u^\nu = 1 \ (ds^2 = g_{\mu\nu}dx^\mu dx^\nu) \text{ and } \delta_{\mu\nu}u^\mu u^\nu = \frac{1}{1 + 2\alpha A_\sigma u^\sigma} \tag{24}$$

In (23), for a while, we remain the term, $e\gamma A_\lambda/c - e\gamma A_\lambda/c$ , intact to show its real meaning. We now introduce the following notation:

$$\bar{e} = e\gamma = \frac{e}{1 + 2\frac{e}{m_0 c^2}A_\mu u^\mu} \tag{25}$$

where $\bar{e}$ is called an *effective charge*. The 4-momentum then takes

$$P_\lambda = \left(m_0 c u_\lambda - \frac{1}{c}\bar{e}A_\lambda\right) + \frac{1}{c}\bar{e}A_\lambda \tag{26}$$

where

$$u_\lambda = g_{\lambda\sigma}u^\sigma = \delta_{\lambda\sigma}(1+2\alpha A_\mu u^\mu)\frac{dx^\sigma}{cdt[(1-\beta^2)(1+2\alpha A_\mu u^\mu)]^{\frac{1}{2}}} = \frac{\dot{x}_\lambda(1+2\alpha A_\mu u^\mu)^{\frac{1}{2}}}{\sqrt{1-\beta^2}}. \quad (27)$$

If one introduces an *effective inertial mass*

$$\bar{m}_0 = m_0(1+2\alpha A_\mu u^\mu)^{\frac{1}{2}}, \quad (28)$$

the space components and time component of (26) can be written as

$$P_i = \left(\frac{\bar{m}_0 V_i}{\sqrt{1-\beta^2}} - \frac{1}{c}\bar{e}A_i\right) + \frac{1}{c}\bar{e}A_i \quad (29)$$

and

$$cP_0 = E = \left(\frac{\bar{m}_0 c^2}{\sqrt{1-\beta^2}} - \bar{e}\varphi\right) + \bar{e}\varphi \quad (30)$$

or

$$E = mc^2 = \frac{\bar{m}_0 c^2}{\sqrt{1-\beta^2}} = \frac{m_0(1+2\alpha A_\mu u^\mu)^{\frac{1}{2}}c^2}{\sqrt{1-\beta^2}}$$

where $\varphi = A_0$: here, $m$ is defined as the inertial mass of a system consisting of a charge and its field including interaction: $m = E/c^2 = \bar{m}_0/\sqrt{1-\beta^2}$. In (29) and (30) the bracket is the part related to a particle and its self-field, and the second term is the part related to the interaction. Thus, the bracket is the momentum of a system consisting of a particle and its self-field. To see this, consider the Newtonian approximations of (29) and (30). In (29), ignoring the terms taking $1/c$ and $1/c^2$ as coefficients, (29) gives

$$\mathbf{P} = m_0\mathbf{V} \quad (31)$$

In (30), if one expands $E$ in powers of $V/c$ and $2eA_\sigma u^\sigma/m_0 c^2$, neglecting the terms of high order ($1/c^2$), the result is

$$\frac{\bar{m}_0 c^2}{\sqrt{1-\beta^2}} \approx m_0 c^2 + \frac{1}{2}m_0 V^2 + e\varphi \text{ and } \bar{e}\varphi \approx e\varphi \quad (32)$$

Thus, the energy leads to

$$E = \left(m_0 c^2 + \frac{1}{2}m_0 V^2 + e\varphi - e\varphi\right) + e\varphi = \left(m_0 c^2 + \frac{1}{2}m_0 V^2\right) + e\varphi \quad (33)$$

In (33), the bracket, according to the starting postulate I, implies the energy of a free particle and its self-field.

Next, consider a charge in the electrostatic field. From (30), the energy of a charge is given by

$$E = E_{(m)} + E_{(f)} = \left[ m_0 c^2 \left(1 + 2\frac{e}{m_0 c^2}\varphi\right)^{1/2} - \bar{e}\varphi \right] + \bar{e}\varphi \tag{34}$$

where $E_{(m)}$ is the energy relative to a particle plus its self field and $E_{(f)}$ the interaction energy: $E_{(f)} = \bar{e}\varphi$ should be viewed as the field energy that a particle and other particles producing external fields own jointly through the interaction. Now, in the bracket of (34), if one expands two terms respectively in powers of $2e\varphi/m_0 c^2$ and neglect terms of order higher than $1/c^2$, we have

$$E_{(m)} \approx m_0 c^2 + \frac{3}{2}\frac{(e\varphi)^2}{m_0 c^2} \tag{35}$$

As already discussed in Sect. 2, the postulate I notes that the total energy of a free particle and its self field equals $m_0 c^2$ (at rest). (35) implies that when a charge is subject to the external field, the energy of a charge and its self field is increased by about $3(e\varphi)^2/2m_0 c^2$ owing to the interaction of the particle with the external field. Hence, in the equation of motion of a charge, the motion of a system (as integrity) consisting of a charge and its self-field is treated: however, as a phrase called "the equation of motion of a charge" traditionally and widely is used in the physics, in the paper, the phrase also is employed intact. In this relation, when getting the equation of motion of a charge, the bracket of (26) should be put on the left side of the equation. On the other hand, (30) gives the relation of energy and inertial mass in a general form. As shown in Sect. 2, the energy formula in SR specifies the relation between the energy of a free particle and mass, while (30), in the presence of an external field, shows a relation between the energy of a particle and inertial mass: in this case, the mass is meant by the measured real mass, i.e., the inertial mass of a system consisting of the particle and its field which contains the interaction with the external field. Consequently, the establishment of CNE enables one to extend and generalize the relation between inertial mass and energy of a system (a free charge plus its self field) in SR into that between inertial mass and energy of a system subject to the external field. Under the weak field condition in the electrostatic field, the relation derived from (30) is expressed as follows:

$$E = \frac{\bar{m}_0 c^2}{\sqrt{1-\beta^2}} \approx \frac{m_0 c^2 \left(1 + 2\frac{e}{m_0 c^2}\varphi\right)^{\frac{1}{2}}}{\sqrt{1-\beta^2}} \approx \frac{m_0 c^2 \left(1 + \frac{e}{m_0 c^2}\varphi\right)}{\sqrt{1-\beta^2}} \approx m_0 c^2 + \frac{1}{2} m_0 V^2 + e\varphi$$

$$= m_0 \left(1 + \frac{e}{m_0 c^2}\varphi + \frac{1}{2}\frac{V^2}{c^2}\right)c^2 \tag{36}$$

(36) shows that, in the presence of the external electrostatic field, the rest mass of a system is varied by $e\varphi/c^2$ and the total mass (including the motion mass) is varied by $e\varphi/c^2 + m_0 V^2/2c^2$. Hence, the interaction energy of the particle with the external field contributes to the variation of the measured mass. Obtained in CNE, the generalized relation between inertial mass and energy of a system provides an extremely important conclusion. In (30), supposing that $2\alpha A_\mu u^\mu < 0$ and $|2\alpha A_\mu u^\mu| = 1$ hold, the numerator of (30) becomes zero and so does the energy. If allowing for the presence of a particle and the energy different from zero, in the nominator of (30), $\beta^2$ must be 1. That is, a particle with nonzero rest mass should be converted into a photon with zero rest mass: for this will be, in detail, discussed in Sect. 4. This implies that (30) specifies the mechanism of annihilation of particles and occurrence of a new particle, which cannot be explained in terms of the Maxwell theory.

Now, deriving the equation of motion of a charge from Eq. (21) yields

$$m_0 c \frac{du_\lambda}{ds} - \frac{1}{c} \bar{e} \frac{dA_\lambda}{ds} = \frac{1}{c} \bar{e} \, F_{\lambda\sigma} u^\sigma \tag{37}$$

where $F_{\lambda\sigma}$ is the electromagnetic field tensor. It is important to note the physical meaning imposed on $-\bar{e} dA_\lambda/cds$, the second term on the left side of (37). This term is derived from $\bar{e}A_\lambda/c$, the second term of the inside of the bracket in (26), which contributes to eliminating the increase of 4-momentum by the interaction of the particle with the external field according to the increase of the effective inertia mass. This is more clearly shown in the formula for energy: see (34-35). While the system of a charge and its field interacts with the external field, contained electromagnetic mass, the effective inertial mass of the system is varied. Thus, an additional inertial force (reaction) arises according to the variation of the effective inertial mass such that changes as little as possible. This inertial force arising from the interaction of the system with the external field is added to 4-acceleration in the form of $(-\bar{e} dA_\lambda/cds)$.

Next, we derive the formula for the energy of a system. From $g_{\lambda\sigma} = \delta_{\lambda\sigma}(1 + 2\alpha A_\mu u^\mu)$ and $g_{\lambda\sigma} g^{\lambda\sigma} = 1$ it follows that

$$g^{\lambda\sigma} = \frac{\delta_{\lambda\sigma}}{1 + 2\alpha A_\mu u^\mu}, \tag{38}$$

$$g^{\lambda\sigma} u_\lambda u_\sigma = \frac{1}{1 + 2\alpha A_\mu u^\mu} (u_0 u_0 - u_i u_i) = 1, \tag{39}$$

and

$$u_0 u_0 - u_i u_i = 1 + 2\alpha A_\mu u^\mu. \tag{40}$$

Using (40) and (26), we obtain

$$(E)^2 - (\mathbf{P})^2 = \bar{m}_0^2 c^4. \tag{41}$$

Arranging (41), the energy of the system becomes

$$E = [\bar{m}_0^2 c^4 + c^2 (\mathbf{P})^2]^{1/2} \tag{42}$$

We now find the approximate formula of Lagrangian to the terms of second-order and discuss the approximate versions of 4-momentum. For this purpose, expanding (9) in series of powers of $2\alpha A_\lambda u^\lambda$, we have the action for the motion of a charge to the terms of second-order:

$$S \approx -m_0 c^2 \int dt \left(1 - \frac{V^2}{c^2}\right)^{\frac{1}{2}} - \frac{e}{c} \int A_\lambda \frac{\dot{x}^\lambda}{(1 + 2\alpha A_\mu \underline{u}^\mu)^{\frac{1}{2}}} dt + m_0 c^2 \frac{1}{2} \int (\alpha A_\lambda \underline{u}^\lambda)^2 \left(1 - \frac{V^2}{c^2}\right)^{\frac{1}{2}} dt$$

$$= -m_0 c \int dt \left(1 - \frac{V^2}{c^2}\right)^{\frac{1}{2}} - \frac{e}{c} \int A_\lambda \frac{\dot{x}^\lambda}{(1 + 2\alpha A_\mu \underline{u}^\mu)^{\frac{1}{2}}} \left(1 - \frac{1}{2} \alpha A_\lambda \underline{u}^\lambda\right) dt$$

$$\approx -m_0 c^2 \int dt \left(1 - \frac{V^2}{c^2}\right)^{\frac{1}{2}} - \frac{e}{c} \int A_\lambda \frac{\dot{x}^\lambda}{1 + \frac{3}{2}\alpha A_\mu \underline{u}^\mu} dt.$$

This can be rewritten in the form of

$$S \approx -m_0 c^2 \int d\underline{s} \, (\delta_{\mu\nu} \underline{u}^\mu \underline{u}^\nu)^{\frac{1}{2}} - \frac{e}{c} \int A_\lambda \frac{\underline{u}^\lambda}{1 + \frac{3}{2}\alpha A_\mu \underline{u}^\mu} d\underline{s}, \qquad (43)$$

where $O(1/c^4)$ is ignored and $d\underline{s}^2 = \delta_{\mu\nu} dx^\mu dx^\nu$ is employed: $\underline{u}^\mu = dx^\mu/d\underline{s}$ is 4-velocity in Mikowski space. From (43) we have the approximate expression of 4-momentum,

$$P_\lambda = \frac{m_0 V_\lambda}{\sqrt{1-\beta^2}} + \frac{1}{c} \frac{e}{\left(1 + \frac{3}{2}\alpha A_\mu \underline{u}^\mu\right)} A_\lambda - \frac{3}{2}\alpha(A_\mu \underline{u}^\mu) \frac{e}{c} A_\lambda \approx \frac{m_0 V_\lambda}{\sqrt{1-\beta^2}} + \frac{1}{c} \frac{e}{(1 + 3\alpha A_\mu \underline{u}^\mu)} A_\lambda$$

by

$$P_\lambda = -\frac{\partial L}{\partial \underline{u}^\lambda}.$$

Noting that $\underline{e} = e/1 + 3\alpha A_\mu \underline{u}^\mu$ ($\underline{e}$ is called a semi-effective charge), the approximate expression of 4-momentum is

$$P_\lambda = \frac{m_0 V_\lambda}{\sqrt{1-\beta^2}} + \frac{1}{c} \underline{e} A_\lambda \qquad (44)$$

If one writes (44) by space and time components of 4-momentum respectively, the result is

$$P_i = \frac{m_0 V_i}{\sqrt{1-\beta^2}} + \frac{\underline{e}}{c} A_i \qquad (45)$$

and

$$E = cP_0 = \frac{m_0 c^2}{\sqrt{1-\beta^2}} + \underline{e}\varphi \qquad (46)$$

Allowing for

$$(E - \underline{e}\varphi)^2 = \left(\frac{m_0 c^2}{\sqrt{1-\beta^2}}\right)^2 \text{ and } \left(\mathbf{P} - \frac{\underline{e}}{c}\mathbf{A}\right)^2 = \left(\frac{m_0 \mathbf{V}}{\sqrt{1-\beta^2}}\right)^2, \qquad (47)$$

we have an approximation expression for Hamiltonian:

$$E = \left[m_0^2 c^4 + c^2 \left(\mathbf{P} - \frac{\underline{e}}{c}\mathbf{A}\right)^2\right]^{\frac{1}{2}} + \underline{e}\varphi. \qquad (48)$$

(48) is essential to get the finite corrections from the S-matrix of quantum nonlinear electrodynamics without invoking renormalization.

Next, we find an approximate expression of the equations of motion of a charge from the Lagrange equation obtained by varying the action (43), whose derivation is the same as in the Maxwell theory:

$$m_0 c \frac{d\underline{u}_\lambda}{d\underline{s}} = \frac{1}{c} \underline{e} F_{\lambda\sigma} \underline{u}^\sigma - \frac{1}{c} A_\lambda \frac{d\underline{e}}{d\underline{s}}, \qquad (49)$$

Eq. (49) holds in the Minkowski space. However, Eq. (49) has a problem. On multiplying the left and right sides of Eq. (49) by 4-velocity $u_\lambda$, the left side becomes zero whereas the right side does not lead to zero. It is due to the fact that Eq. (49) is the approximate expression: such an analogy is found in the equation of motion of a charge taking into account radiation damping in the Maxwell theory; see p 226 of Ref. 22. When considering the interaction of a charge with the field, the approximate condition characterizing whether the field is strong or weak and whether the interaction is made at the regime of high energy or of low energy is defined and estimated by the interaction energy scale of a particle and fields, $b_0 = |2\alpha A_\lambda u^\lambda| \approx |2e\varphi/m_0 c^2|$; see (13). From now on, we note that $|2e\varphi/m_0 c^2| \ll 1$ is the weak field condition or low energy condition, and the case in which $|2e\varphi/m_0 c^2|$ is not so smaller than 1 is the semi-strong field condition or semi-high energy condition. The strong field condition corresponds to $|2e\varphi/m_0 c^2| \approx |2eA_\lambda u^\lambda/m_0 c^2| \cong 1$. When considering Eq. (49) under the semi-strong energy condition, the term taking it as a coefficient, $|2eA_\lambda u^\lambda|/m_0 c^3$ smaller than $|2eA_\lambda u^\lambda/m_0 c^2|$ is neglected. To see this, we transform Eq. (49) as follows:

$$m_0 c \frac{du_\lambda}{d\underline{s}} = \frac{1}{c} e \underline{F}_{\lambda\sigma} \underline{u}^\sigma - 2 \frac{e^2}{m_0 c^3} A_\lambda \frac{d(A_\mu \underline{u}^\mu)}{d\underline{s}} (1 + 3\alpha A_\lambda \underline{u}^\lambda)^{-2}. \tag{50}$$

In Eq. (50), the third term is neglected:

$$2 \frac{e^2}{m_0 c^3} A_\lambda \frac{d(A_\mu \underline{u}^\mu)}{d\underline{s}} \approx 2 \frac{e^2}{m_0 c^3} A_\lambda (\text{grad}\varphi \cdot \mathbf{V}) \sim 2 \frac{e^2}{m_0 c^3} A_\lambda |\mathbf{V}||\text{grad}\varphi| \approx 0$$

Thus, on considering Eq. (50) under the semi-strong energy condition, the second term of the right side of Eq. (50) vanishes in relevant approximation. Consequently, Eq. (50) leads to

$$m_0 c \frac{du_\lambda}{d\underline{s}} = \frac{1}{c} e \underline{F}_{\lambda\sigma} \underline{u}^\sigma \tag{51}$$

When multiplying the left and right sides of (51) by 4-velocity $\underline{u}_\lambda$, both sides of the equation become zero. Eq. (51) is the equation of motion of a charge, which holds under the semi-strong field condition or at the regime of semi-high energy.

## 4. Experimental verification for theoretical results under strong field condition

In order to demonstrate the validity of the action (8), it is very important to show that the theoretical results derived from the action (8) are in agreement with experiments, while the results cannot be explained in the terms of the Maxwell theory. Unfortunately, within an extremely small spatial region satisfying the conditions of the strong or semi-strong field as quantum effects notably appear, a complete description for the motion of a charge under these conditions cannot be given. However, it should be emphasized that CNE gives a description of some essential characteristics of the actual mechanism that underlies the motion.

### 4.1. Pair annihilation of an electron and positron, and production of photons

First of all, we consider the pair annihilation of electron and positron and the production of photons.

In this case, as the functions characterizing physical quantities in the action (9) and Eq. (37), due to the disappearance of particles, have discontinuous and singular points at $r = r_0$, they can not be treated with differential equations. Therefore, in terms of the conservation law of energy of the system (electron-positron), we study this phenomenon. According to (30), the energy of the system is expressed as follows:

$$E = \left[ \frac{m_0 c^2 \left(1 + 2\frac{e_{(-)}}{m_0 c^2} A_\mu^{(+)} u_1^\mu \right)^{\frac{1}{2}}}{\sqrt{1 - \frac{v_1^2}{c^2}}} - \bar{e}_{(-)} \varphi^{(+)} \right] + \frac{1}{2} \bar{e}_{(-)} \varphi^{(+)} + \frac{1}{2} \bar{e}_{(+)} \varphi^{(-)}$$

$$+ \left[ \frac{m_0 c^2 \left(1 + 2\frac{e_{(+)}}{m_0 c^2} A_\mu^{(-)} u_2^\mu \right)^{\frac{1}{2}}}{\sqrt{1 - \frac{v_2^2}{c^2}}} - \bar{e}_{(+)} \varphi^{(-)} \right] \quad (52)$$

where $e_{(-)}$ and $e_{(+)}$ are the charges of electron and positron, $\bar{e}_{(-)}$ and $\bar{e}_{(+)}$ their effective charges, $A_\mu^{(-)}$ and $A_\mu^{(+)}$ 4-field vector produced by electron and positron, $A_\mu = A_\mu^{ex} + A_\mu^{in}$ ($A_\mu^{ex}$; external field, $A_\mu^{in}$; self-field), $u_1^\mu$ the 4-velocity of an electron, $u_2^\mu$ the 4-velocity of a positron: the form of $(\bar{e}_{(-)} \varphi^{(+)}/2 + \bar{e}_{(+)} \varphi^{(-)}/2)$ is rooted in the formula for the total energy of a system consisting of charges and fields produced by them in the Maxwell theory.

The detailed arguments for $A_\mu$ will be given in Sect. 8. For the present study, if writing the results only, we have

$$A_\lambda = -\frac{1}{c} \cdot \frac{\bar{e} V_\lambda \left(t - \frac{r}{c}\right)}{r} = -\frac{1}{c} \cdot \frac{e V_\lambda \left(t - \frac{r}{c}\right)}{r[1 + 2\alpha(A_\mu^{in}(r_a) + A_\mu^{ex}(r_a)) u^\lambda]} \quad (53)$$

where $A_\lambda$ is 4-field vector produced by the effective charge $\bar{e} = e/[1 + 2\alpha(A_\mu^{in}(r_a) + A_\mu^{ex}(r_a)) u^\lambda]$, (as shown in (25), the constant charge $e$ in the Maxwell theory is replaced by the effective charge $\bar{e}$) and $A_\mu^{in}(r_a)$ and $A_\mu^{ex}(r_a)$ are the self-field and the external field at the point $r_a$ which a particle is located: $A_\mu^{in} = 0$, in the case a free charge, holds and in the presence of radiation reaction, $\varphi^{in} \approx \lim_{R \to 0} -(e/6c^3) \partial^3 R^2/\partial t^3$ and $\mathbf{A}^{in} \approx -e\dot{\mathbf{V}}/c^2$, and so

$$\mathbf{E}^{in} \approx -\text{grad}\varphi^{in} - \frac{1}{c} \frac{\partial \mathbf{A}^{in}}{\partial t} = \frac{2\ddot{\mathbf{V}}}{3c^3} \quad (54)$$

In (52), allowing for

$$e_{(-)} A_\mu^{(+)} = e_{(+)} A_\mu^{(-)},$$

$$\frac{2e_{(-)} A_\mu^{(+)} u_1^\mu}{m_0 c^2} \approx \frac{2e_{(-)} \varphi^{(+)}}{m_0 c^2} \approx -2 \frac{e^2}{m_0 c^2} \frac{1}{r} + O\left(\frac{1}{c^4}\right),$$

and

$$\bar{e}_{(-)}\varphi^{(+)} = \bar{e}_{(+)}\varphi^{(-)} = -\frac{1}{r} \cdot \frac{|e|}{\left(1 - 2\frac{e^2}{m_0 c^2}\frac{1}{r} + O\left(\frac{1}{c^4}\right)\right)} \cdot \frac{|e|}{\left(1 - 2\frac{e^2}{m_0 c^2}\frac{1}{r} + O\left(\frac{1}{c^4}\right)\right)}, \quad (55)$$

and then if neglecting the terms of high order more than $r_0 = 2e^2/m_0 c^2$ (electron radius; $r_0 \approx 10^{-13}$ cm), (52) arrives at

$$E = \frac{m_0 c^2 (1 - r_0/r)^{1/2}}{\sqrt{1 - v_1^2/c^2}} + \frac{m_0 c^2 (1 - r_0/r)^{1/2}}{\sqrt{1 - v_2^2/c^2}} + \frac{e^2}{r(1 - r_0/r)^2} \quad (56)$$

When the electron and positron reach $r = r_0$, the denominator of the term relevant to the interaction energy (the third term of the right side of (56)) becomes zero. In order for the energy to have a finite value, the charge $e$ should be zero, namely, at $r = r_0$ charges vanish. When considering the terms relevant to the particle in (56), at $r = r_0$, the numerator becomes zero. In order for the energy of a particle to have a finite value, not zero, $v_1^2/c^2 = v_2^2/c^2 = 1$ should hold in the denominator, that is, the particle should have the velocity of light, $v = c$, followed by zero rest mass. This fact follows from the requirement of SR that a free particle with nonzero rest mass cannot reach the velocity of light. Actually at $r = r_0$, $e = 0$ must hold, and then, owing to the absence of interaction of particles, the particle becomes a free particle. From this, it follows that the electron-positron annihilation gives birth to the production of two photons, which agrees with experiment. Consequently, we have

$$\left\{\frac{m_0 c^2 (1 - r_0/r)^{\frac{1}{2}}}{\sqrt{1 - v_1^2/c^2}} + \frac{m_0 c^2 (1 - r_0/r)^{\frac{1}{2}}}{\sqrt{1 - v_2^2/c^2}} + \frac{e^2}{r(1 - r_0/r)^2}\right\}\bigg|_{r=r_0} = 2\hbar\omega \quad (57)$$

If the former theory holds, one naturally leads to a wrong conclusion. In the Maxwell theory, the energy of an electron-positron system is expressed as follows:

$$E = \frac{m_0 c^2}{\sqrt{1 - v_1^2/c^2}} + \frac{m_0 c^2}{\sqrt{1 - v_2^2/c^2}} - \frac{e^2}{r} \quad (58)$$

When the electron and positron arrive at a specific point $r = 0$, for finiteness of the energy, the charge $e$ in the third term of the right side of (58) should be zero, namely, at $r = 0$ charges vanish. But, in this case, two "neutral particles" with the same mass as an electron or positron may possibly occur after the interaction of an electron and positron. It is because of the conclusion that only electric charges but not masses vanish at $r = 0$ after the electron-positron interaction.

### 4.2. Electron capture by a proton

Next, we consider an electron capture by a proton. Referring to (56), the energy of an electron-proton system is expressed as

$$E = \frac{m_0 c^2 \left(1 - \frac{r_0}{r}\right)^{\frac{1}{2}}}{\sqrt{1 - \frac{v_1^2}{c^2}}} + \frac{M_0 c^2 \left(1 - \frac{r_0^p}{r}\right)^{\frac{1}{2}}}{\sqrt{1 - \frac{v_2^2}{c^2}}} + \frac{1}{r} \cdot \frac{e}{\left(1 - \frac{r_0}{r}\right)} \cdot \frac{e}{\left(1 - \frac{r_0^p}{r}\right)} \approx$$

$$\approx \frac{m_0 c^2 \left(1 - \frac{r_0}{r}\right)^{\frac{1}{2}}}{\sqrt{1 - \frac{v_1^2}{c^2}}} + \frac{M_0 c^2 \left(1 - \frac{r_0^p}{r}\right)^{\frac{1}{2}}}{\sqrt{1 - \frac{v_2^2}{c^2}}} + \frac{1}{r} \cdot \frac{e^2}{\left(1 - \frac{r_0}{r}\right)} \tag{59}$$

where $r_0^p$ is the proton radius ($r_0^p = e^2/M_0 c^2$), $M_0$ is the rest mass of a proton ($M_0 = 1836 \, m_0$), and $r_0^p \ll r_0$. When the electron reaches $r = r_0$, what will result in? As far as the electron reaches $r = r_0$, a lifetime of the electron ends and it follows from (59) that $e = 0$, $\bar{m}_0 = 0$, and $v_1^2/c^2 = 0$: $\bar{m}_0 = m_0(1 - r_0/r)^{1/2}$.

*Note*: we can assume that in (59) the electron is not captured by the proton but converted into a photon, i.e. $e = 0$, $m_0 = 0$, and $v_1^2/c^2 = 1$ ($v_1 = c$). But this conversion cannot be permitted, because it stands against the conservation law of charge. A proton retains the lifetime as far as its antiparticle does not arrive at the space point which $r = r_0^p$ holds. Thus, the electron is annihilated to be captured by the proton and then the proton-electron system is converted into a neutron devoid of charge. From this, we have the following conservation formula of the energy of the system:

$$\left\{ \frac{m_0 c^2 (1 - r_0/r)^{1/2}}{\sqrt{1 - v_1^2/c^2}} + \frac{M_0 c^2 (1 - r_0^p/r)^{1/2}}{\sqrt{1 - v_2^2/c^2}} + \frac{1}{r} \cdot \frac{e^2}{(1 - r_0/r)} \right\}\bigg|_{r=r_0} = \frac{M_0^n c^2}{\sqrt{1 - v_3^2/c^2}} \tag{60}$$

where $M_0^n$ is the mass of the neutron and $v_3$ the velocity of a neutron. This was also verified experimentally. However, from the former theory does not follow the possibility of electron capture by a proton. According to the Maxwell theory, a proton and an electron can approach infinitely close irrespective of the difference between their masses, leading to a wrong result that at $r = 0$, the charge of electron and proton becomes zero respectively and then two neutral particles ("neutral electron" and neutron) generate. Of cause, in this case, it is not concluded that the mass of the electron becomes zero. The annihilation of a charged particle is meant by the simultaneous disappearance of charge and mass. But the former theory does not yield this conclusion. If so, what is the reason? It is due to the fact that CNE inevitably provides the strong interference (the strong dependence of the effective charge and effective inertial mass on the external field) in the neighborhood of $r = r_0$, which is absent in the Maxwell theory. That is, the Maxwell theory gives no solution to the mechanism of the annihilation and occurrence of particles.

The annihilation and occurrence of particles are singular physical effects arising from strong fields at close distances between particles (particle radius). In a field theory, experimental verifications of physical effects manifested in strong fields are of decisive significance to confirm the validity of the theory. It is since under strong fields are most clearly manifested fundamental effects that show intrinsic and immanent characteristics peculiar to the theory. For example, in Einstein's GR, recent discoveries of Black Hole (Black Star) producing a highly strong gravitational field in the vicinity of a gravitational singular point made a great contribution to demonstrate the validity of GR. Likewise, the solution to the mechanism of the annihilation and production of particles also is of decisive significance to verify the validity of CNE. Concerning the argument mentioned above, some comments may be presented that the annihilation and occurrence of particles can be possibly studied only with the aid of quantum electrodynamics. But, this is a wrong view. In the Maxwell theory, the energy formula of a free particle $E = m_0 c^2/\sqrt{1 - v^2/c^2}$ specifies the existence of a photon: when a particle arrives at the velocity of the light, for the finiteness of the energy a special particle (so-called photon) with zero rest mass must be present. However, the Maxwell theory gives no description of the mechanism of photon production because this phenomenon arises under the strong field condition whereas the application of the theory is confined to the regime of a weak field. In contrast with the Maxwell theory, CNE provides a good description of this phenomenon, which is a major advance in classical electrodynamics. Of course, CNE gives no probability analysis of the annihilation and production of particles such as the scattering cross-section. This study is just imposed on quantum electrodynamics.

## 5. Generalized conservation law of charge and limit of applicability of gauge symmetry

In modern physics, the most commonly accepted consensus is that gauge symmetry is a well-defined principle, which is closely connected with the conservation law of charge, a basic law of nature. A breaking of gauge symmetry, perhaps, may give a great impact on modern physics. In this section, we show that CNE, with a broken gauge symmetry, enables one to have a more universal and inclusive understanding of the conservation law of charge, and then discuss the limit of applicability of gauge symmetry. At first, we briefly touch on the close relationship between the gauge symmetry and the conservation law of charge, which has been phrased in the Maxwell theory; see p 78 of Ref. 22.

In Maxwell theory, the action for the motion of a charge is expressed as follows:

$$S = -\sum \int mc\,ds - \frac{1}{c^2} \int A_\lambda j^\lambda \, d\Omega \tag{61}$$

where $j^\lambda$ is 4-current density: $j^\lambda = (c\rho, \mathbf{j})$, where $\rho$ is charge density and $\mathbf{j}$ is the current density vector. On replacing $A_\lambda$ by $A_\lambda - (\partial f/\partial x^\lambda)$, the integral

$$\frac{1}{c^2} \int j^\lambda \frac{\partial f}{\partial x^\lambda} d\Omega \tag{62}$$

is added to the second term of (61). It is explicitly the conservation of charge, as expressed by $\partial j^\lambda/\partial x^\lambda = 0$, that enables us to write the integral as 4-divergence $\partial(fj^\lambda)/\partial x^\lambda$, after which, using Gauss theorem, the integral over a volume of four dimensions is transformed into that over a hypersurface: on varying the action, this integral drops out, and thus, has no effect on the equation of motion. On the other hand, as far as the conservation formula of a charge in differential form ($\partial j^\lambda/\partial x^\lambda = 0$) holds, the charge $e$ of individual particles is always conserved as a constant at arbitrary coordinates of space-time, and so the conservation of the total charge of the system is expressed by an arithmetic sum of respective charge conservations. This implies that $\partial j^\lambda/\partial x^\lambda = 0$ is a necessary condition for holding the gauge symmetry, but not just enough one. Supposing that the interaction Lagrangian includes a nonlinear term, for example, $\alpha(A_\lambda j^\lambda)^2$ as in the action (43), LSP is broken and so the gauge symmetry is also invalid. Therefore, only when the conservation law of charge in differential form ($\partial j^\lambda/\partial x^\lambda = 0$) and LSP are valid simultaneously, the gauge symmetry holds. This shows that $\partial j^\lambda/\partial x^\lambda = 0$ and LSP are the necessary and enough conditions for satisfying the gauge symmetry. But, in CNE, $\partial j^\lambda/\partial x^\lambda = 0$ and LSP, in general, does not hold, and so, the gauge symmetry is broken as well. Hence, problems arise whether or not the conservation law of charge is present and if the law is present, how it is expressed. In this context, we provide a description of generalized charge conservation that holds in CNE. The total charge of a closed system is always conserved, which has been accepted as an axiom in classical electrodynamics. The conservation of the total charge of the system is expressed by the equation of continuity in integral form. In curvilinear coordinates, we have

$$\frac{\partial}{\partial t}\int \rho\sqrt{-g}dV = -\oint \mathbf{j}\sqrt{-g}d\mathbf{f} \tag{63}$$

where $\rho$ is the charge density and $\mathbf{j}$ the current density ($\mathbf{j} = \rho\mathbf{v}$, $\mathbf{v}$; the velocity of charge ): for point charges, the 4-current vector is given by

$$j^\lambda = \sum_a \frac{e_a c \delta(\mathbf{r}-\mathbf{r}_a)}{\sqrt{-g}}\frac{dx^\lambda}{dx^0} \tag{64}$$

Substituting (64) in (63), we can easily confirm that the equation of continuity in integral form in curvilinear coordinates (63) is the same as in a Minkowski space. By applying Gauss' theorem to (63), we have the law of charge conservation in differential form:

$$\frac{1}{\sqrt{-g}}\frac{\partial\sqrt{-g}j^\lambda}{\partial x^\lambda} = 0 \tag{65}$$

where (65) is expressed by a covariant derivative. Substituting (64) in (65), it is shown that the conservation law in differential form in curvilinear coordinates is also the same as in the Minkowski space. For particles with discontinuous distribution of charge, (65) leads to

$$\sum_a \left[\text{div} e_a \mathbf{v}\delta(\mathbf{r}-\mathbf{r}_a) + \frac{\partial e_a \delta(\mathbf{r}-\mathbf{r}_a)}{\partial t}\right] = 0 \tag{66}$$

where $e_a$ is the charge of an arbitrary particle at the spatial coordinate vector $\mathbf{r}_a$ which the particle is located, Eq. (66) shows that, during the motion of a particle, the charge of the particle $e_a$ is always conserved as a constant. However, in CNE, the conservation law (66) in differential form is not generally derived from the conservation law (63) in integral form. To see this, take an example that, after a particle beam consisting of electrons and positrons is incident on a certain volume, through the surface bounding the volume, and then the particle pairs (electron-positron) of some part of the beam bring about the annihilation of the particles and the production of photons, the remaining particles go to the outside of the volume. In the presence of the annihilation of a particle with charge, special emphasis should be given to the fact that functions characterizing charge and current densities are not continuous and smooth, since some particles with charge vanish at some specific spatial coordinates. As well known, only when a function is continuous and smooth, Gauss' theorem holds. Thus, as seen in (65) and (66), the law of charge conservation in differential form is not derived from the charge conservation in integral form, and therefore, under the strong field condition that the annihilation and production of particles arise, the gauge symmetry is broken and only the conservation law in integral form holds. Consequently, *under the strong field condition, only the total charge of a system before and after the interaction of particles but not the individual charge is conserved, while the gauge symmetry and LSP are broken simultaneously.*

Under the semi-strong field condition, as the annihilation and production of particles do not arise, the conservation law in integral form, as well as that in differential form holds together. In such a system, not only the total charge of system but also the individual charges of particles are conserved, while the gauge symmetry and LSP are broken simultaneously, as shown in (43) and (49). Under the weak field condition, as shown in the Maxwell theory, since LSP and the conservation law in differential form hold together, the gauge symmetry is valid, and then the individual charges of particles, as well as the total charge of system are conserved. This indicates the limit of applicability of the gauge symmetry and furthermore of the Maxwell theory. In mathematical form, the generalized conservation law is shown intensively in Eq. (36). From Eq. (36), we have

$$\frac{d}{dt}\left(\frac{e}{1 + 2\frac{e}{m_0 c^2}A_\mu u^\mu}\right) \neq 0 \quad (67)$$

Eq. (67) presents the generalized conservation law of charge that holds in the strong and semi-strong fields, as well as weak field, the law which is manifested in some special forms, according to different energy scales of interactions. To see this, replacing the charge by the charge density in (67), (67) arrive at

$$\frac{d\bar{e}\delta(\mathbf{r}-\mathbf{r}_0)}{cdt} = \frac{d}{cdt}\left(\frac{e\ \delta(\mathbf{r}-\mathbf{r}_0)}{1 + 2\frac{e}{m_0 c^2}A_\mu u^\mu}\right) =$$

$$= \frac{1}{c\left(1 + 2\frac{e}{m_0 c^2}A_\mu u^\mu\right)}\frac{\partial j^\lambda}{\partial x^\lambda} + e\delta(\mathbf{r}-\mathbf{r}_0)\frac{d}{cdt}\left(\frac{1}{1 + 2\frac{e}{m_0 c^2}A_\mu u^\mu}\right) \neq 0 \quad (68)$$

where $d/dt = (\mathbf{v}\cdot\text{grad}, \partial/\partial t\ )$ and $j^\lambda = (ce\delta(\mathbf{r}-\mathbf{r}_0), e\mathbf{v}\delta(\mathbf{r}-\mathbf{r}_0))$.

A. Under the weak field condition $2eA_\mu u^\mu/m_0 c^2$ is ignored and a charge is always conserved as a constant:

$$\frac{d\bar{e}\delta(\mathbf{r}-\mathbf{r}_0)}{cdt} = 0 \quad \text{and} \quad j^\lambda/\partial x^\lambda = 0. \quad (69)$$

B. Under the semi-strong field condition, a charge also is conserved as a constant: $j^\lambda/\partial x^\lambda = 0$.

$$\frac{d\bar{e}\delta(\mathbf{r}-\mathbf{r}_0)}{cdt} = e\delta(\mathbf{r}-\mathbf{r}_0)\frac{d}{cdt}\left(\frac{1}{1 + 2\frac{e}{m_0 c^2}A_\mu u^\mu}\right) \quad (70)$$

C. Under the strong field condition, $|2eA_\lambda u^\lambda/m_0 c^2| = 1$, which the annihilation of charges arises, $(1 + 2eA_\mu u^\mu/m_0 c^2) = 0$ holds and so, in order for (67) to have meaning, the charge $e$ must vanish. In this case, Eq. (68) does not hold owing to a singular point. Thus, $\partial j^\lambda/\partial x^\lambda = 0$ is not valid and only the conser- vation law of charge in integral form holds.

As mentioned above, CNE provides the conservation law of charge that contains a more universal and inclusive meaning as compared to the Maxwell theory.

## 6. Energy-momentum tensor and field equations

This section focuses on finding the field equations. The action for motion of a particle, (9) does not obey LSP. In CNE, the field equations should be also non-linear equations that are not subject to LSP and gauge symmetry. Unfortunately, in the paper, because of the mathematical poverty of authors, nonlinear differential field equations have been not found in the most general form. Instead, we will get approximate field equations valid within some limit of applicability.

Using the action (9) for motion of a particle and applying $\delta$-function to it, the total action for particle and field is given by

$$S = \int L\sqrt{-g}d\Omega = \int d\Omega\sqrt{-g}(L_m + L_f) = \int d\Omega\sqrt{-g}\left[-\frac{m_0 c}{\sqrt{-g}}(g_{\mu\nu}\dot{x}^\mu\dot{x}^\nu)^{\frac{1}{2}}\delta(\mathbf{r}-\mathbf{r}_0) - \frac{1}{16\pi c}L_f\right] \tag{71}$$

where $L_m$ is the Lagrangian for motion of a particle, $L_f$ is the Lagrangian for field, and $(g_{\mu\nu}\dot{x}^\mu\dot{x}^\nu)^{\frac{1}{2}} = c(1-V^2/c^2)^{\frac{1}{2}}(1+2\alpha A_\lambda u^\lambda)^{\frac{1}{2}}$. As a due fact $L_f$ should constitute a function that satisfies the conservation law of total energy-momentum of particle and field. Therefore, to find the condition that $L_f$ should obey, we first find the conservation law of total energy-momentum of particle and field using the variational principle.

For the variation $\delta g_{\mu\nu}$ in $g_{\mu\nu}$, using Gauss' theorem and setting $\delta g_{\mu\nu} = 0$ at the limits of integration, we find $\delta S$ in the following form (see p. 291 of Ref. 22):

$$\delta S = \int \delta(L\sqrt{-g})d\Omega = \frac{1}{c}\int\left\{\frac{\partial L\sqrt{-g}}{\partial g^{\mu\nu}} - \frac{\partial}{\partial x^\lambda}\frac{\partial L\sqrt{-g}}{\partial\frac{\partial g^{\mu\nu}}{\partial x^\lambda}}\right\}\delta g^{\mu\nu}d\Omega = 0 \tag{72}$$

Here we introduce the notation

$$\frac{1}{2}\sqrt{-g}T_{\mu\nu} = \frac{\partial L\sqrt{-g}}{\partial g^{\mu\nu}} - \frac{\partial}{\partial x^\lambda}\frac{\partial L\sqrt{-g}}{\partial\frac{\partial g^{\mu\nu}}{\partial x^\lambda}} \tag{73}$$

and then, using the detailed derivation showed in p. 291 of Ref. 22, we have

$$\delta S = -\frac{1}{c}\int\frac{1}{\sqrt{-g}}\frac{\partial\sqrt{-g}T_\mu^\nu}{\partial x^\nu}\xi^\mu\sqrt{-g}\,d\Omega = -\frac{1}{c}\int T_{\mu;\nu}^\nu\xi^\mu\sqrt{-g}\,d\Omega \tag{74}$$

where $x'^\mu = x^\mu + \delta x^\mu = x^\mu + \xi^\mu$, ($\xi^\mu$; small quantities). Because of the arbitrariness of the $\xi^\mu$, it follows that

$$T_{\mu;\nu}^\nu = \frac{1}{\sqrt{-g}}\frac{\partial\sqrt{-g}T_\mu^\nu}{\partial x^\nu} = 0 \tag{75}$$

From (73) we obtain the energy-momentum tensor of particle and field,

$$T_{\mu\nu}{}^{(m)} = 2\frac{1}{\sqrt{-g}}\frac{\delta(L_m\sqrt{-g})}{\delta g^{\mu\nu}} = \frac{1}{\sqrt{-g}}m_0 c u_\mu u_\nu \frac{ds}{dt}\delta(\mathbf{r}-\mathbf{r}_0),$$

or

$$T_\mu^{\nu(m)} = \frac{1}{\sqrt{-g}} m_0 c u_\mu u^\nu \frac{ds}{dt} \delta(\mathbf{r} - \mathbf{r}_0), \tag{76}$$

where $T_{\mu\nu}{}^{(m)}$ and $T_\mu^{\nu(m)}$ are the energy-momentum tensors of particle, and its covariant derivative is given by

$$\frac{1}{\sqrt{-g}} \frac{\partial \sqrt{-g} T_\mu^{\nu(m)}}{\partial x^\nu} = \frac{1}{\sqrt{-g}} m_0 c \frac{du_\mu}{dt}. \tag{77}$$

The energy-momentum tensor of field $T_{\mu\nu}{}^{(f)}$ is obtained from

$$T_{\mu\nu}{}^{(f)} = 2 \frac{1}{\sqrt{-g}} \frac{\delta(L_f \sqrt{-g})}{\delta g^{\mu\nu}}. \tag{78}$$

Thus, the conservation law of energy-momentum of particle and field is expressed as

$$\frac{\partial \sqrt{-g} T_\mu^{\nu(m)}}{\partial x^\nu} + \frac{\partial \sqrt{-g} T_\mu^{\nu(f)}}{\partial x^\nu} = 0. \tag{79}$$

(77) - (79) must give rise to Eq. (36):

$$\frac{\partial \sqrt{-g} T_\mu^{\nu(m)}}{\partial x^\nu} + \frac{\partial \sqrt{-g} T_\mu^{\nu(f)}}{\partial x^\nu} = m_0 c \frac{du_\mu}{ds} - \left(\frac{1}{c} \bar{e} F_{\lambda\sigma} u^\sigma + \frac{1}{c} \bar{e} \frac{dA_\lambda}{ds}\right) = 0 \tag{80}$$

From (77) and (80), $L_f$ should constitute a function that satisfies the following equation:

$$\frac{\partial \sqrt{-g} T_\mu^{\nu(f)}}{\partial x^\nu} = 2 \frac{\partial}{\partial x^\nu} \left(\frac{g^{\lambda\nu}}{\sqrt{-g}} \frac{\delta(L_f \sqrt{-g})}{\delta g^{\mu\lambda}}\right) =$$

$$= -\left(\frac{1}{c} \bar{e} F_{\lambda\sigma} u^\sigma + \frac{1}{c} \bar{e} \frac{dA_\lambda}{ds}\right) \tag{81}$$

In appearance, Eq. (81) contains an additional term, $\bar{e} dA_\lambda/cds$, compared to the Maxwell theory. Unfortunately, we do not have a complete $L_f$. In this context, $L_f$ is obtained by an approximate approach. For this purpose, we rewrite Eq. (51) which holds under the semi-strong field condition:

$$m_0 c \frac{du_\lambda}{d\underline{s}} = \frac{1}{c} \underline{e} F_{\lambda\sigma} \underline{u}^\sigma \tag{82}$$

The term $\bar{e} dA_\lambda/cds$ in Eq. (81) that $L_f$ should satisfy is absent in Eq. (82).

Now, find $L_f$ approximately such that the conservation law of energy-momentum of particle and field (79) yields Eq. (82). Such an approximate $L_f$ must necessarily have the form of $F_{\mu\nu} F^{\mu\nu}$. To see this, we first find the field equations in an approximate version.

$$S = \int L\sqrt{-g} d\Omega = \int d\Omega \sqrt{-g} (L_m + L_f) = \int d\Omega \sqrt{-g} \left[-\frac{m_0 c^2}{\sqrt{-g}} (g_{\mu\nu} \dot{x}^\mu \dot{x}^\nu)^{\frac{1}{2}} \delta(\mathbf{r} - \mathbf{r}_0) - \frac{1}{16\pi c} F_{\mu\nu} F^{\mu\nu}\right]$$

(83)

where $d\Omega = cdtdxdydz$ and $g_{\mu\nu} = \delta_{\mu\nu}(1 + 2\alpha\delta^{\lambda\sigma}A_\lambda u_\sigma)$ : $u_\sigma = dx_\sigma/cdt(1 - V^2/c^2)^{\frac{1}{2}}(1 + 2\alpha A_\mu u^\mu)^{1/2}$. In the action (83), the first term constitutes a complete expression of the action for motion of a particle, whereas the second term is an approximate version of the action for field. If one uses

$$F^{\mu\nu} = g^{\mu\lambda}F_\lambda^\nu = g^{\mu\lambda}g^{\nu\sigma}F_{\lambda\sigma} = -\frac{1}{\sqrt{-g}}F_{\mu\nu} \tag{84}$$

and regards the motion of a particle as being given, from the variational principle the field equation of the approximate version is derived:

$$\delta S = \int \left\{ \frac{\partial \sqrt{-g}L_m}{\partial A_\mu}\delta A_\mu + \frac{1}{8\pi c}F_{\mu\nu}\delta F_{\mu\nu} \right\} d\Omega \tag{85}$$

Referring to p.79 of Ref. 22, we have

$$\delta S = \int \frac{1}{c}\left[-\frac{1}{c}\frac{e\delta(\mathbf{r} - \mathbf{r}_0)V_\mu}{(1 + 2\alpha A_\lambda(\mathbf{r}_0)u^\lambda)} + \frac{1}{4\pi}\frac{\partial F_{\mu\nu}}{\partial x^\nu}\right]\delta A_\mu d\Omega - \frac{1}{4\pi}\int F_{\mu\nu}\,\delta A_\mu dS_\nu, \tag{86}$$

where the second term leads to zero. Therefore, (86) arrives at

$$\frac{\partial F_{\mu\nu}}{\partial x^\nu} = \frac{4\pi}{c}\bar{e}\delta(\mathbf{r} - \mathbf{r}_0)V_\mu = \frac{4\pi}{c}\frac{e\delta(\mathbf{r} - \mathbf{r}_0)}{1 + 2\alpha A_\lambda(\mathbf{r}_0)u^\lambda}V_\mu, \tag{87}$$

where $A_\lambda(\mathbf{r}_0) = A_\lambda^{in}(\mathbf{r}_0) + A_\lambda^{ex}(\mathbf{r}_0)$. Taking 4-divergence on both sides of Eq. (87), the left side leads to zero, whereas the right side yields

$$\frac{d(\bar{e}\delta(\mathbf{r} - \mathbf{r}_0))}{dt} = 2\frac{4\pi e^2\delta(\mathbf{r} - \mathbf{r}_0)}{m_0 c^3}\frac{d(A_\lambda(\mathbf{r}_0)u^\lambda)}{dt}[1 + 2\alpha A_\lambda(\mathbf{r}_0)u^\lambda]^{-2} \tag{88}$$

Under the semi-strong field condition, (88) is neglected, while Eq. (87) holds: see (50).

In Eq. (87), since the identity $\partial^2 F^{\mu\nu}/\partial x^\mu \partial x^\nu = 0$ holds, an additional condition, the Lorentz condition is obtained:

$$\frac{\partial A^\nu}{\partial x^\nu} = 0 \tag{89}$$

After imposing the Lorentz condition on the potentials, we get

$$\Box \mathbf{A} = -\frac{4\pi}{c} \cdot \frac{e\mathbf{V}}{1 + 2\alpha(A_\lambda^{in}(\mathbf{r}_0) + A_\lambda^{ex}(\mathbf{r}_0))u^\lambda}\delta(\mathbf{r} - \mathbf{r}_0) \tag{90}$$

and

$$\Box \varphi = -4\pi \cdot \frac{e}{1 + 2\alpha(A_\lambda^{in}(\mathbf{r}_0) + A_\lambda^{ex}(\mathbf{r}_0))u^\lambda}\delta(\mathbf{r} - \mathbf{r}_0) \tag{91}$$

where $A_\lambda^{ex}(\mathbf{r}_0)$ is the external field potentials and $A_\lambda^{in}(\mathbf{r}_0)$ is the self-field potentials: $A_\lambda^{ex}(\mathbf{r}_0)$ and $A_\lambda^{in}(\mathbf{r}_0)$ are the field potentials at the spatial coordinate vector $\mathbf{r}_0$ which a charge is located. Thus, the field potentials produced by the charge are dependent on the external field and self-field. This implies that the interactions between charges are dependent or interfered each other. This is just an interference effect peculiar to CNE, which cannot be described in terms of the Maxwell theory. The interference effect arises due to the dependence of the effective charge on external field, which is absent in the linear theory, that is, the Maxwell theory. As already studied in Sect. 4, the annihilation of particles and production of new particles is the most typical phenomenon arising from the interference effect at high energy regimes.

Next, it is shown that the Lagrangian for field $L_f$ is in agreement with the conservation law of energy-momentum within some approximation. The action for field is as follows:

$$S_f = \frac{1}{16\pi c}\int F_{\mu\nu}F^{\mu\nu}\sqrt{-g}\,d\Omega = \frac{1}{16\pi c}\int F_{\mu\nu}F_{\lambda\sigma}g^{\mu\lambda}g^{\nu\sigma}\sqrt{-g}\,d\Omega \tag{92}$$

where $S_f$ is the action for field. From (92), (73), and (78), the energy-momentum tensor of field is obtained (see p. 293 of Ref. 22):

$$T^{(f)}_{\mu\nu} = -\frac{1}{4\pi}\left(F_{\mu\lambda}F_\nu^{\ \lambda} - \frac{1}{4}F_{\lambda\sigma}F^{\lambda\sigma}g_{\mu\nu}\right) \tag{93}$$

or

$$T^{\nu(f)}_\mu = -\frac{1}{4\pi}\left(F_{\mu\lambda}F^{\nu\lambda} - \frac{1}{4}F_{\lambda\sigma}F^{\lambda\sigma}\delta_\mu^\nu\right), \tag{94}$$

where $T^{\nu(f)}_\mu$ is the energy-momentum tensor of field and $\delta_\mu^\nu = g^{\nu\lambda}g_{\mu\lambda} = \delta^{\nu\lambda}\delta_{\mu\lambda}$: $\delta_{\mu\lambda}$ is the Minkowski metric. Referring to p. 86 of Ref. 22 for the detailed derivation, we find

$$\frac{\partial\sqrt{-g}\,T^{\nu(f)}_\mu}{\partial x^\nu} = -\frac{1}{c}\bar{e}F_{\mu\nu}V^\nu\delta(\mathbf{r}-\mathbf{r}_0) \tag{95}$$

with $\sqrt{-g}\,F^{\mu\nu} = -F_{\mu\nu}$; see (84). If one substitutes Eq. (95) and Eq. (77) in Eq. (79) and arranges it, Eq. (79) leads to

$$\frac{\partial\sqrt{-g}\,T^{\nu(m)}_\mu}{\partial x^\nu} + \frac{\partial\sqrt{-g}\,T^{\nu(f)}_\mu}{\partial x^\nu} = m_0 c\frac{du_\mu}{dt} - \frac{1}{c}\bar{e}F_{\mu\nu}\frac{dx^\nu}{dt} \tag{96}$$

Allowing for $2\alpha A_\lambda u^\lambda < 1$, from relevant approximation, the first term of Eq. (96) leads to

$$m_0 c\frac{du_\mu}{ds} = m_0 c\frac{d(g_{\mu\lambda}u^\lambda)}{ds} \approx m_0 c\left[\frac{d\underline{u}_\mu}{d\underline{s}} + \frac{e}{m_0 c^3}\frac{d(A_\lambda \underline{u}^\lambda)}{dt}\frac{1}{\gamma}\underline{u}_\mu\right] \tag{97}$$

where $\underline{u}_\mu = dx_\mu/d\underline{s}$ is the 4-velocity in Minkowski space, $d\underline{s} = cdt\sqrt{1-\beta^2}$, and $\gamma = \sqrt{1-\beta^2}(1+2\alpha A_\lambda\underline{u}^\lambda)$; see (28) for $u^\lambda$ and $u_\mu$, and here $(e/m_0 c^3)d(A_\lambda u^\lambda)/dt \approx (e/m_0 c^3)d(A_\lambda\underline{u}^\lambda)/dt + O(1/c^5)$ is employed. On the other hand, the second term of Eq. (96) leads to

$$\frac{1}{c}\bar{e}F_{\mu\nu}u^\nu = \frac{1}{c}\frac{e}{(1+2\alpha A_\lambda u^\lambda)^{\frac{3}{2}}}F_{\mu\nu}\underline{u}^\nu \approx \frac{1}{c}\frac{e}{(1+3\alpha A_\lambda\underline{u}^\lambda)}F_{\mu\nu}\underline{u}^\nu + O(1/c^5) \tag{98}$$

From the definition for the semi-strong field condition, if neglecting the second term of the inside of the bracket in (97) and $O(1/c^5)$ in (98), since Eq. (82) holds, (97) and (98) become

$$m_0 c \frac{d\underline{u_\lambda}}{d\underline{s}} - \frac{1}{c}\underline{e}F_{\lambda\sigma}\underline{u}^\sigma \approx 0 \tag{99}$$

and

$$m_0 c \frac{d\underline{u_\lambda}}{dt} - \frac{1}{c}\underline{e}F_{\lambda\sigma}V^\sigma \approx 0 \tag{100}$$

From (96) and (100), it follows

$$\frac{\partial \sqrt{-g}T_\mu^{\nu(m)}}{\partial x^\nu} + \frac{\partial \sqrt{-g}T_\mu^{\nu(f)}}{\partial x^\nu} \approx 0 \tag{101}$$

Consequently, under the semi-strong field condition, the equation of the motion of a charge (82) and field equation (87) are subjected to the conservation law of energy-momentum.

## 7. KR space and normalization of main functions characterizing particle-field

In this section we propose KR space (a new kind of non-Euclidean space) that underlies CNE. For a reasonable description of the electromagnetic field theory of point particles, why do we must introduce so complicated non-Euclidean space-KR space? The whole history of physics proves that space and time are a form of existence of matter and the geometry of space-time is determined by the essential content of matter. As well known in the history of physics, in Newton's classical mechanics were introduced the relativity principle of Galileo and three-dimensional Euclidean geometry. But later, the starting idea of Einstein, especially the principle of invariance of light velocity, revolutionized our thinking of matter, and then gave birth to four-dimensional space-time. And, in 1916, from the principle of equivalence in Einstein's GR was introduced the non-Euclidean four-dimensional Riemann space as a real space. Likewise, the starting postulate in which the total energy of a particle plus its field should be $mc^2$, and the correspondence principle naturally gives birth to KR space, and the action is formulated on it.

### 7.1. KR space

It was well known that Euclidean space corresponds to any point (or infinitesimal space region) of Riemann space $g_{\mu\nu}(x)$:

$$g_{\mu\nu} = g_{\mu\nu}(x) \text{ and } g'_{\mu\nu} = g_{\mu\nu}(x = x_0 = constant) = \eta_{\mu\nu}. \tag{102}$$

where $g_{\mu\nu}$ is the Riemannian metric tensor and $g'_{\mu\nu} = \eta_{\mu\nu}$ is Euclidean metric tensor. On the other hand, Riemannian space corresponds to any point of Finsler space $g_{\mu\nu}(x, \dot{x})$:

$$g_{\mu\nu} = g_{\mu\nu}(x, \dot{x}) \quad and \quad g_{\mu\nu}^R = g_{\mu\nu}(x, \dot{x} = constant) = g'_{\mu\nu}(x) \tag{103}$$

where $g_{\mu\nu}$ is the Finsler metric tensor and $g_{\mu\nu}^R$ is the Riemann metric tensor. In a word, Riemannian space includes Euclidean space as a special form, and Finsler space includes Riemannian space as a special form. But a non-Euclidean space discussed in this paper is neither Riemannian space nor Finsler space. The action for particle and field considered in Sect. 3 and Sect. 6 were defined on a new non-Euclidian space whose metric tensor is

$$g_{\mu\nu} = \delta_{\mu\nu}(1 + 2\alpha A_\lambda u^\lambda) \quad \text{or} \quad g^{\mu\nu} = \delta^{\mu\nu}\frac{1}{1 + 2\alpha A_\lambda u^\lambda} \tag{104}$$

where $g_{\mu\nu} = g_{\mu\nu}(x, u)$; see the action (8). The field function $A_\lambda$ to be studied in Sect. 8 has the following form:

$$A_\lambda \sim \frac{\underline{A}_\lambda}{(1 + 2\alpha A_\sigma u^\sigma)} = \underline{A}_\lambda g^{\mu\nu}\delta_{\mu\nu}/4 = A_\lambda(x, \dot{x}, g), \tag{105}$$

where $\underline{A}_\lambda$ is the field function defined in the Minkowski space (obtained in the Maxwell theory), that is, $\underline{A}_\lambda \sim e\dot{x}_\lambda/r$ and $\delta_{\mu\nu}\delta^{\mu\nu} = 4$ is employed. On the other hand, in (104) $u^\lambda$ are expressed as

$$u^\lambda = \frac{dx^\lambda}{dt(g_{\mu\nu}\dot{x}^\mu\dot{x}^\nu)^{1/2}} = u^\lambda(\dot{x}, g). \tag{106}$$

Consequently, substituting (105) and (106) in (104), it follows that

$$g_{\lambda\sigma} = g_{\lambda\sigma}(x, \dot{x}, g_{\mu\nu}), \tag{107}$$

where we define $g_{\mu\nu}$ as a variable metric and $g_{\lambda\sigma}$ as a function metric. As shown in (107), the function metric $g_{\lambda\sigma}$ has a new character dependent on the metric tensor itself (the variable metric $g_{\mu\nu}$) in addition to coordinate and velocity. Because $A_\alpha$ and $u^\alpha$ enter the variable metric $g_{\mu\nu}$ in (107) and so $g_{\mu\nu}$ is also dependent on the variable-metric $g$, the function metric $g_{\lambda\sigma}$ has the character of absolute implicit function which cannot be, by any means, transformed into an explicit function. In this connection, we define a new non-Euclidean space called "KR space". If one fixes the variable metric $g_{\mu\nu}$ in the function metric $g_{\lambda\sigma}(x, \dot{x}, g_{\mu\nu})$ by the Euclidean constant metric tensor $\eta_{\mu\nu}$, the function metric $g_{\lambda\sigma}$ results in the Finsler metric tensor $g'_{\lambda\sigma}(x, \dot{x})$:

$$g_{\lambda\sigma}(x, \dot{x}, g_{\mu\nu} = \eta_{\mu\nu}) = g'_{\lambda\sigma}(x, \dot{x}). \tag{108}$$

Consequently, KR space is meant by the space whose metric tensor, as a function of metric tensor itself, owns the implicit characteristic of function, and has a feature that Finsler space corresponds to any point $(x, \dot{x})$ of the space. Namely, when $g_{\mu\nu} = \delta_{\mu\nu}$, Eq. (108) arrives at

$$g'_{\lambda\sigma} = \delta_{\lambda\sigma}(1 + 2\alpha\underline{A}_\mu\underline{u}^\mu), \tag{109}$$

with

$$\underline{u}^\mu = \underline{u}^\mu(\dot{x}) \text{ and } \underline{A}_\mu = \underline{A}_\mu(x, \dot{x}),$$

where $\underline{A}_\mu(x, \dot{x})$ and $\underline{u}^\mu(\dot{x})$ are the four-dimensional functions in the Maxwell theory, as defined in Minkowski space and $\delta_{\mu\nu}$ is Minkowski metric. Because of the character of the absolute implicit function of the metric tensor on KR space, it is impossible to argue its realistic meaning necessary for measurement. To solve this problem, in the first step, we introduce an idea of the normalization of the metric tensor. The variable metric $g_{\mu\nu}$ can be fixed by an arbitrary constant metric tensor in every point $(x, \dot{x})$ of KR space, and Finsler space corresponding to $(x, \dot{x})$ is therefore not uniquely determined. We define it as the normalization of metric tensor that one obtains the metric tensor of Finsler space, through fixing the variable metric $g_{\mu\nu}$ by Minkowski metric tensor $\delta_{\mu\nu}$ in any point $(x, \dot{x})$ of KR space. From this definition, we get

$$\bar{g}_{\lambda\sigma} = g_{\lambda\sigma}(x, \dot{x}, g|_\delta) = g_{\lambda\sigma}(x, \dot{x}), \tag{110}$$

where $\bar{g}_{\lambda\sigma}$ is the normalized metric tensor and $g|_\delta$ denotes $g_{\mu\nu} = \delta_{\mu\nu}$. The metric tensor obtained from Eq. (110) is obviously the Finsler metric tensor and is uniquely given at every point $(x, \dot{x})$: KR space has not been thoroughly studied mathematically and hence in this paper we also do not offer a rigorous mathematical definition and analysis for KR space. The normalization of a metric is the key to argue the normalization of implicit function in the next sections.

**7.2. Normalization rules of implicit function and its physical meaning**

Here, we provide the description of normalization rules and the physical meaning of the implicit function, which are the key to solve divergence problems arising from classical and quantum electrodynamics. As already seen above, functions characterizing field and particle on KR space, because of the character of the implicit function of the metric tensor, also become implicit ones. In the paper, the implicit function has the following general form:

$$F = F(x, \dot{x}, F'(x, \dot{x}, g), f'(x, \dot{x}, g)) \tag{111}$$

or

$$F = F(x, \dot{x}, g)$$

with $g = g(x, \dot{x}, F', f')$, where $F$ and $F'(x, \dot{x}, g)$ have the same form, while the prime is used for the distinction between function and variable, and $f'(x, \dot{x}, g)$ is all the functions dependent on the metric tensor $g$. For example, in (105), $F'(x, \dot{x}, g)$ is the field potential $A_\lambda(x, \dot{x}, g)$, and $f'(x, \dot{x}, g)$ is the four-dimensional velocity $u_\lambda(\dot{x}, g)$ on KR space. the general form of implicit function, in the paper, is expressed as $F = F(x, \dot{x}, g)$ and $g = g(x, \dot{x}, F', f')$.

*Note*: In a rigorous mathematical viewpoint, implicit function is not expressed as (111). In mathematics, the implicit function is defined as $y = f(x)$ satisfying the equation $G(x, y) = 0$. In a strict sense, the form of (111) denotes an equation for getting the implicit function, and the implicit function should be represented as $F = f(x, \dot{x})$ satisfying (111). However, in the paper, the function $F$, under a relevant approximate condition, is transformed into the explicit function (defined in the Maxwell theory) and furthermore, as shown in this section, through the normalization of $F$, the implicit function $F$ is transformed into the normalized function $\bar{F}$ (explicit function). To specify such a process clearly, we call the function $F$ the implicit function and exhibit the general form of the implicit function $F$ as (111).

Generally, if the coordinates of space-time are given, the value of a function is uniquely defined. But on KR space, owing to the specific implicit character of function, even though the coordinates are given, it is impossible to find the value of the function determined uniquely. In this regard, we should first present how to give real meaning to implicit function to get measurable physical quantities. The real physical meaning of the implicit function $F$ is determined by the normalized function $\bar{F}$. In this relation, we define the second starting postulate as a qualification equivalent to the first postulate proposed in Sect. 2. **Postulate II**: *on KR space, every physical quantity is expressed as an implicit function, and functions (normalized functions) that have real physical meaning are determined as explicit functions uniquely corresponding to Finsler space according to the rules of normalization.*

The normalization of implicit function is related to the normalization of the metric tensor. It is because a function includes the metric tensor as a variable and the metric tensor also includes the function itself as a variable, from which follows the implicit character of the function.

Now, we define the rules of normalization as follows.

Rule 1: as for an implicit function $F$ characterizing field or the interaction of a particle and field on KR space, we define it as the normalization of the implicit function $F$ that one fixes $g_{\mu\nu}$ (or $g$), the variable metric of $F'(x,\dot{x},g)$ and $f'(x,\dot{x},g)$ (producing the implicit character of function) entered $F(x,\dot{x},F'(x,\dot{x},g),f'(x,\dot{x},g))$ in (111) by the Minkowski metric tensor $\delta_{\mu\nu}$, and obtain $\bar{F}$ corresponding to Finsler space, called "normalized function":

$$\bar{F} = F(x,\dot{x},F'(x,\dot{x},g|_\delta),f'(x,\dot{x},g|_\delta)) = F(x,\dot{x}). \tag{112}$$

where $\bar{F}$ is the normalized function. In the case of $F = F(x,\dot{x},g)$ and $g = g(x,\dot{x},F',f')$, (112) yields

$$\bar{F} = F(x,\dot{x},\bar{g}_{\mu\nu}). \tag{113}$$

where the normalization of the metric tensor proceeds according to Eq. (110). and then (113) is derived easily from (112). This rule plays a key role in eliminating an infinity of physical quantities and getting finite quantities.

Rule 2: the 4-velocity of a particle is expressed as follows (see (106)):

$$u^\lambda = u^\lambda(\dot{x},g), \tag{114}$$

where the variable metric $g$ is given by

$$g_{\mu\nu} = g_{\mu\nu}(x,\dot{x},g). \tag{115}$$

As for an implicit function $u^\lambda$, we define the normalization of implicit function $u^\lambda$ as follows:

$$\bar{u}^\lambda = u^\lambda(\dot{x},\bar{g}) \tag{116}$$

with

$$\bar{g}_{\lambda\sigma} = \delta_{\lambda\sigma}\big(1 + 2\alpha\bar{A}_\mu u^\mu(\dot{x},g|_\delta)\big) = \delta_{\lambda\sigma}\big(1 + 2\alpha\bar{A}_\mu u^\mu(\dot{x})\big),$$

where $\bar{A}_\mu$ is the already normalized function according to rule 1 and is used as it is.

Rule 3: for a function $\varepsilon(x,F)$, the normalized function $\bar{\varepsilon}$ is as follows:

$$\bar{\varepsilon} = \bar{\varepsilon}(x, \bar{F}), \qquad (117)$$

where $\bar{F}$ is the normalized function of $F$.

Rule 4: when a function $\varepsilon$ consists of the sum of $\varepsilon_1, \varepsilon_2, \cdots, \varepsilon_n$, the normalization of $\varepsilon$ is the same as the sum of the normalizations of individual terms:

$$\bar{\varepsilon} = \bar{\varepsilon}_1 + \cdots + \bar{\varepsilon}_i + \cdots + \bar{\varepsilon}_k + \cdots + \bar{\varepsilon}_n. \qquad (118)$$

Rule 5: if $\varepsilon$ is the expansion of series and $\varepsilon_k (k > i)$ includes $\varepsilon_i (i \geq 1)$ in (118), the normalization of $\varepsilon$ is performed in serial order from $\varepsilon_1$ to $\varepsilon_n$, and as for normalization of $\varepsilon_k (k > i)$, the already normalized $\bar{\varepsilon}_i$ is used.

The idea and the rules of normalization are essential to establish CNE. If so, what is the true essence and significance of the introduction of implicit function? Why should physical quantities be expressed as implicit functions and what is the true meaning of implicit function? If one first says the conclusion, what physical quantity is expressed as implicit function is owing to a feedback of particle-field. As well known, a particle acts on itself through the field produced by the particle. In other words, particle-field has a character of feedback.

Particle → field (the field is created by a particle)

Field → particle (the field acts on the particle)

The result which involves such a character is revealed as just implicit function. But when one introduces this feedback to Maxwell-Lorentz covariant theory, that is, the linear electrodynamics, it was well known that unavoidable divergences occur [14-17]. As we shall see in the next sections, this problem of divergence is very easily solved by using the normalization rules following from that, on KR space the character of feedback of particle-field is reflected in the metric tensor and functions characterizing physical quantities, and so every physical quantity is expressed as implicit function. Then, why does the feedback in field theory appears at all? The modern "systems theory" has already clarified that feedback is an important aspect of the nature of matter. As far as an external action acts on matter, some effect surely occurs. Here the external action constitutes the cause giving birth to the effect. For example, in Newtonian mechanics, the external action is meant by an external force and in biology, it is a change of the external environment. According to "systems theory", the occurred effect is applied again to the external action giving birth to it. Furthermore, feedback brings about a reaction, which is rooted in a self-maintenance or self-resistance character of matter which is going to reduce the effect by the external action to a possible extent. For example, in Newtonian mechanics, the reaction occurs owing to the mechanical inertia of an object and in electrodynamics there exists the radiation reaction by the feedback of radiation wave arising due to the electromagnetic inertia. Furthermore, in biology, whenever the surrounding temperature increases, an endothermic reaction arises in an organism, and whenever it decreases, an exothermic reaction occurs to keep the temperature of the organism constant. As a result, feedback is based on the nature of matter and implicit function is the mathematical manifestation of feedback. From this follows the conclusion that the manifestation of implicit function is not artificially introduced but comes naturally from the nature of matter. In this relation, it is an admirable fact that Mandel Sachs presented, already long ago, the following good prediction; one more feature of matter that must be introduced in a unified field theory is the inertial manifestation of matter, that is, the resistance of matter to a change of its state of rest or constant motion and it will be key to achieving a unified field theory [19]. This is a very important conclusion. If one asks a question about why feedback and normalization rules must be introduced, we have to ask the question to matter! It is due to an inevitable result coming from the immanent and intrinsic nature of matter, which controls the interaction of matter with self-field: when the field produced by the matter comes to be a variable of the field function by feedback, the matter does not receive the field function as a variable right away but as a version of the function transformed by the normalization rule: in the automatic control engineering (imitating the self-control of organism), when the output signal of a system is returned to the input, the returned signal is transformed according to the self-control of the system. It does not need any theoretical demonstration and so the description of implicit function and feedback comes to be starting postulate II as a physical axiom, which yields theoretical results without renormalization in agreement with experiments. We also can find such an analogy in quantum mechanics. When studying the motion of matter, a problem one faces is to be impossible to describe the motion mathematically as it is in reality: e.g. a particle with some size and form in classical mechanics and an electron with wave property. In this relation, the modeling approach is well-defined in physics. For example, in quantum mechanics, one introduced the wave model (absent in reality) and complex wave function (devoid of measurable and realistic meaning) and defined the real physical meaning of wave function as $|\psi|^2$ to describe the characteristics of matter-wave. Similarly, eliminating infinite quantities and finding the characteristics of particle-field without any contradiction bring about the introduction of implicit function as an inevitable consequence of feedback character in the model of a point particle, and the real physical meaning of an implicit function $F$ is defined by the normalized implicit function $\bar{F}$. The matter has wave-particle dualism. As a result, the descriptions of wave and particle properties have an interesting symmetry with some correspondences listed in Table I: it is due to commonness in the modeling approach. Thus, just as the introduction of complex wave function was indispensable to describe mathematically the wave nature of the matter in quantum theory, the implicit function was introduced for a consistent description of the particle nature and the normalization of implicit function was defined for clarification of its physical meaning. As will be discussed in the next sections, the finiteness of physical quantities is obtained, based on this idea of normalization.

Table I. Symmetries in the descriptions of wave property and particle property.

|  | **wave property** | **particle property** |
|---|---|---|
| **model** | wave (no exist realistically) | point particle (no exist realistically) |
| **mathematical description** | complex wave function (unmeasurable quantity) | implicit function on KR space (unmeasurable quantity) |
| **physical meaning** | be defined as $|\Psi|^2$ | be defined as normalized function |

## 8. Main results of classical nonlinear electrodynamics (CNE)

In physics, the Maxwell theory has been recognized as a nearly complete and unique classical theory for the electromagnetic field: despite the presence of so many papers which indicated difficulties in the Maxwell theory [1-17]. Experimental laws which underlie the Maxwell theory were clearly discovered and argued within a weak field. The weak field is meant by a field in which the interaction energy of particles is much less than the rest energy, $m_0 c^2$: $U/m_0 c^2 \ll 1$ ($U$; interaction energy). In this case, $r \gg r_0$ ($r$; interaction radius, $r_0$; around electron radius, $\sim 10^{-13}\ cm$) is satisfied and the interaction between particle and field is dealt only in a much farther distance than the electron radius, $e^2/m_0 c^2 \approx 10^{-13} cm$. Simply speaking, in the Maxwell theory, the electromagnetic fields produced by charged macro-matter and macro-electric current were considered. But there is no special reason that experiments discovered within the weak field should always be in agreement with those in the strong field. In the view of sensible science, new characteristics that cannot be found in the weak field are inevitably exhibited in the strong field. For example, at an extremely near specific distance of the interaction between charges, the pair annihilation of particle and antiparticle arise and so within such a small spatial region Coulomb law is invalid owing to disappearance of charges. However, it is impossible to perform all experiments in the whole area comprising not only the weak field but also the strong field. Thus, that one uses not only an inductive method but also a deductive method based on main principles of physics to establish a more generalized theory, is essential for the building of a consistent theory. In this regard, the main difficulties of the Maxwell theory are rooted in concluding that the experimental results obtained in the weak field are universal truths satisfied in the whole region comprising the weak field and strong field, and then in making no more advance into the establishment of a more generalized and consistent theory. We have already argued in Sect. 4 and Sect. 5 that the theoretical results are in good agreement with experiments obtained under the strong field condition, which cannot be explained by the Maxwell theory. In this case, we assumed that the functions of field potential are already given and gave no detailed description or derivations. In this section, we describe the solution of field equations and radiation reaction, and based upon it, discuss the relation between the inertial mass and total energy of particle plus field.

### 8.1. Equation of electrostatic field and variation of Coulomb law

From the characteristics of the electrostatic field (a constant field not dependent on time), only the scalar potential is discussed. From Eq. (94), we get the equation of the scalar potential:

$$\nabla^2 \varphi = -4\pi \frac{e}{1 + 2\alpha \varphi(r_0, g) u^0} \delta(\mathbf{r} - \mathbf{r}_0), \qquad (119)$$

where $\alpha = e/m_0 c^2$. Note: As shown in (120) and (104), since $\varphi = \varphi(R, g) = e/R\left(g^{\mu\nu}\delta_{\mu\nu}/4\right)$ generally holds, the scalar potential $\varphi$ entered the right side of Eq. (119) is also expressed as $\varphi = \varphi(r_0, g)$, while, if one fixes $g^{\mu\nu}$ by $\delta_{\mu\nu}$, the scalar potential $\varphi$ leads to the Coulomb potential in the Maxwell theory.

Eq. (119), from the character of the electrostatic field, satisfies the condition $d\bar{e}/dt = 0$, and from this equation, we obtain the solution for the potential of the electrostatic field. As the effective charge density on the right side of Eq. (119) is given at the spatial coordinate vector $\mathbf{r}_0$, we have the solution of the above equation by the well-known method:

$$\varphi = \int \frac{e}{R(1 + 2\alpha\varphi(r_0, g)u^0)} \delta(\mathbf{r} - \mathbf{r}_0) dV = \frac{e}{R(1 + 2\alpha\varphi(r_0, g)u^0)} + C, \quad (120)$$

where $R = |\mathbf{r} - \mathbf{r}_0|$, and $\varphi(r_0, g)$ is the potential at the spatial coordinate vector $\mathbf{r}_0$ where the charge is located. $\varphi(r_0, g)$ consists of two components, i.e., $\varphi^{ex}(r_0, g)$ (the potential of the external field) and $\varphi^{in}(r_0, g)$ (the potential of the self-field):

$$\varphi(r_0, g) = \varphi^{ex}(r_0, g) + \varphi^{in}(r_0, g). \quad (121)$$

When finding the solution (120) of Eq. (119), we used the following formula:

$$\nabla^2 \frac{f(r_0)}{R} = -4\pi f(r_0)\delta(\mathbf{r} - \mathbf{r}_0), \quad (122)$$

where $f(r_0) = e/(1 + 2\alpha\varphi(r_0, g|_{r=r_0})u^0)$, and $\nabla^2$ acts on the variable $R$ only. From (120) and (121), we have

$$\varphi(R) = \frac{e}{R[(1 + 2\alpha(\varphi^{ex}(r_0, g) + \varphi^{in}(r_0, g))u^0)]} + C. \quad (123)$$

Now, we find $\varphi^{in}(r_0, g)$ from (123). The field acting on the charge itself is define as

$$\varphi^{in}(R) = \lim_{R \to 0} \varphi(R) = \lim_{R \to 0} \frac{e}{R[(1 + 2\alpha(\varphi^{ex}(r_0, g) + \varphi^{in}(r_0, g))u^0)]} + C, \quad (124)$$

where $\left[\left(1 + 2\alpha\left(\varphi^{ex}(r_0, g) + \varphi^{in}(r_0, g)\right)u^0\right)\right]^{-1}$ is the component of the metric tensor of KR space and from (104) holds

$$\left[\left(1 + 2\alpha\left(\varphi^{ex}(r_0, g) + \varphi^{in}(r_0, g)\right)u^0\right)\right]^{-1} = g^{\mu\nu}\delta_{\mu\nu}/4 \quad (125)$$

As $g = g(x, \dot{x}, g)$ holds (see (107)), $\varphi^{in}(r_0, g)$ is the absolute implicit function. The normalization of implicit function $\varphi^{in}$, subject to the normalization rule 1 (see (112) and (113)) yields $\bar{\varphi}^{in} = \varphi^{in}(r_0, \bar{g})$:

$$\bar{\varphi}^{in}(r_0) = \lim_{R \to 0} \varphi(R) = \lim_{R \to 0} \frac{e}{R}\delta_{\mu\nu}\bar{g}^{\mu\nu}/4$$

$$= \lim_{R \to 0} \frac{e}{R\left[\left(1 + 2\alpha(\varphi^{ex}(r_0, g = \delta) + \varphi^{in}(r_0, g = \delta))u^0(\dot{x}, g|_\delta)\right)\right]} + C$$

$$= \lim_{R \to 0} \frac{e}{R\left[\left(1 + 2\alpha\left(\underline{\varphi}^{ex}(r_0) + \underline{\varphi}^{in}(r_0)\right)\underline{u}^0\right)\right]} + C, \quad (126)$$

where $\underline{\varphi}^{in}$ and $\underline{\varphi}^{ex}$ are the potentials obtained in the Maxwell theory, and $\underline{u}^0 = u^0(\dot{x}, g|_\delta) = 1$: $\underline{u}^0$ is the time component of the 4-velocity in Minkowski space, and $\beta = 0$ holds in the electrostatic field. Thus, (126) leads to

$$\bar{\varphi}^{in} = \lim_{R \to 0} \frac{e}{R\left[1 + 2\alpha\left(\frac{e}{R} + \underline{\varphi}^{ex}(r_0)\right)\right]} + C = \frac{m_0 c^2}{2e} + C, \qquad (127)$$

where the potential of the external field $\underline{\varphi}^{ex}(r_0)$ is given automatically by the initial conditions. On the other hand, according to the correspondence principle, the action for the motion of a free charge should be the same as that in the special theory of relativity (SR). Therefore, (127) arrives at

$$\bar{\varphi}^{in} = \frac{m_0 c^2}{2e} + C = 0 \quad \text{and} \quad C = -\frac{m_0 c^2}{2e}. \qquad (128)$$

According to the normalization rule 3 ($\bar{\varepsilon} = \bar{\varepsilon}(x, \bar{F})$), the electrostatic potential $\bar{\varphi}$ can be written as

$$\begin{cases} \bar{\varphi} = \dfrac{e}{R\left[\left(1 + 2\alpha\left(\overline{\varphi}^{ex}(r_0) + \bar{\varphi}^{in}(r_0)\right)\right)\right]} + C \\ \bar{\varphi}^{in}(r_0) = 0 \\ C = 0 \end{cases}, \qquad (129)$$

where $\bar{\varphi}^{ex}(r_0)$ is the normalized potential of the external field and is given by the initial condition, which is the same as $\underline{\varphi}^{ex}(r_0)$ in some approximation. In Eq. (129), the constant $C$, as the field vanishes at infinity, should become zero. It is remarkable that, in the paper, the constant in the electrostatic potential is not an arbitrary value as in the Maxwell theory but determined uniquely under some physical condition. The formula (129) presents the modified version of Coulomb potential.

(129) yields important conclusions.

First, As can be easily understood in (129), the electrostatic potential depends on the external field $\varphi^{ex}$ and so the theory is not in agreement with the LSP. Therefore, the interference effects are expected to arise remarkably within the semi-strong field condition: $r/r_0$ is not enough smaller than 1.

Second, from (129) it follows that an electron and a positron cannot approach infinitely close to each other. The interaction energy of electron-positron can be written as follows:

$$U = -\frac{e^2}{r\left(1 - 2\dfrac{e^2}{m_0 c^2} \cdot \dfrac{1}{r}\right)}. \qquad (130)$$

When the electron and positron approach around the electron radius $r = 2e^2/m_0 c^2 \approx 10^{-13}$ cm, the energy of interaction, $U$, diverges. In order for the potential energy of interaction to possess a finite value, $e^2$ in the numerator should vanish. From this, it is concluded that the electron radius $2e^2/m_0 c^2$ is a critical distance which the electron and positron can approach and as far as the electron and positron approach this distance, the annihilation of an electron and positron and the production of new particles should arise: for this was already discussed in Sect. 4. But, from Coulomb law in the Maxwell theory is not drawn such a conclusion. Thus, modifying the Coulomb law such that the mechanism of the pair annihilation of particle and antiparticle is derived is inevitable.

Third, the potential formula (129) helps us to solve the divergence problem of the total energy of a particle and its self-field, and get the relation between the total energy of particle plus field and the inertia mass, defined in the first starting postulate I; see Subsect. 8.3.

## 8.2. Radiation reaction and natural elimination of divergence by normalization

Here, we find the potentials of self-field which contribute to the radiation reaction produced by a moving charge and obtain the radiation reaction force that gives a good solution to the divergence problem in the potential of radiation reaction without invoking renormalization. For this purpose, we rewrite the field equations discussed in Sect. 6:

$$\Box A_\lambda = -\frac{4\pi}{c} \bar{e} V_\lambda \delta(\mathbf{r} - \mathbf{r}_0). \tag{131}$$

we find the solution of Eq. (131) according to the well-known method:

$$\varphi = \frac{\bar{e}(t - R/c)}{R} \quad \text{and} \quad \mathbf{A} = \frac{1}{c} \cdot \frac{\bar{e}\mathbf{V}(t - R/c)}{R}. \tag{132}$$

When finding (132), (122) was taken into account. In (132) the effective charge, a source of the field depends on the interaction term of particle and field, $(A_\lambda u^\lambda)$. This is the key to solving the divergence problem in radiation reaction. In the Maxwell theory when finding the radiation reaction force by means of renormalization procedure, the field potentials and Lorentz force cannot be used right away because the field potentials are singular in the vicinity of a point charge: in this case, the most widely used approach is to get the radiation reaction force through the renormalization procedure proceeding from the energy-momentum tensor or Green's function; see references [6,15,16]. This derivation requires complicated mathematical computations. But, in the paper, the field potentials can be derived right away and from it, the radiation reaction force is easily drawn. On the other hand, in the Bopp-Podolsky electrodynamics which modified the Maxwell theory, the final result for radiation reaction force gave rise to serious drawbacks: the appearance of a physically invalid term $\mu q^2 \omega/2$ ($\mu$; constant, $q$; charge quantity, $\omega$; acceleration of a charge) in the first term of the series expansion, the auxiliary field with a negative value, the additional term depending on the higher derivatives of the acceleration $\omega^i$, etc. [23]. In the paper, the problems mentioned above are not presented.

When taking into account the radiation reaction, from (9) the action for the motion of a charge can be represented as follows:

$$S = -m_0 c^2 \int dt \left(1 - \frac{V^2}{c^2}\right)^{\frac{1}{2}} (1 + 2\alpha A_\lambda u^\lambda)^{\frac{1}{2}} = -m_0 c \int dt \left(1 - \frac{V^2}{c^2}\right)^{\frac{1}{2}} \left[1 + 2\alpha (A_\lambda^{in} + A_\lambda^{ex}) u^\lambda\right]^{\frac{1}{2}}$$

$$= -m_0 c \int dt \left(1 - \frac{V^2}{c^2}\right)^{\frac{1}{2}} \left[1 + 2\alpha (A_\lambda^{in} \dot{x}^\lambda + A_\lambda^{ex} \dot{x}^\lambda) \gamma\right]^{\frac{1}{2}} \tag{133}$$

where $A_\lambda = A_\lambda^{in} + A_\lambda^{ex}$, $\gamma = (1 - V^2/c^2)^{-\frac{1}{2}} (g^{\mu\nu} \delta_{\mu\nu}/4)^{1/2}$: see (125), and $A_\lambda^{in}$, as a finite quantity, is a very small quantity with high orders of $1/c^2$ and $1/c^3$. On the other hand, the solutions (132) of the field equations are given by

$$\mathbf{A} = -\frac{1}{c} \cdot \frac{e\mathbf{V}\left(t - \frac{R}{c}\right)}{R\left[1 + 2\alpha \left(A_{\lambda(1)}^{in}(r_0, g) + A_\lambda^{ex}(r_0, g)\right) u^\lambda\right]} + \mathbf{C}_1 \tag{134}$$

and

$$\varphi = \frac{e\left(t - \frac{R}{c}\right)}{R\left[1 + 2\alpha\left(A^{in}_{\lambda(1)}(r_0, g) + A^{ex}_{\lambda}(r_0, g)\right)u^\lambda\right]} + C_0, \tag{135}$$

where $A^{ex}_\lambda$ is the potential of the external field and $A^{in}_{\lambda(1)}$ is the first term (the divergent term in the Maxwell theory) of the expansion of potential acting on the particle itself in powers of $R/c$. The terms more than the first order were ignored in our consideration because they give quantities of order higher than $1/c^4$.

In case of $V \ll c$, the expansion of potential gives

$$\varphi = b\frac{e}{R} - b\frac{1}{c} \cdot \frac{\partial e}{\partial t} + b\frac{e}{2c^2} \cdot \frac{\partial^2 R}{\partial t^2} - b\frac{e}{6c^3} \cdot \frac{\partial^3 R^2}{\partial t^3} = \varphi_{(1)} + \varphi_{(2)} + \varphi_{(3)} + \varphi_{(4)} \tag{136}$$

and

$$\mathbf{A} = b\frac{1}{c} \cdot \frac{e\mathbf{V}}{R} - b\frac{e}{c^2} \cdot \frac{\partial \mathbf{V}}{\partial t} = \mathbf{A}_{(1)} + \mathbf{A}_{(2)} \tag{137}$$

with

$$b = \frac{1}{1 + 2\alpha(A^{in}_{\lambda(1)} + A^{ex}_\lambda)u^\lambda}$$

where $\varphi_{(1)}$, $\varphi_{(2)}$, $\varphi_{(3)}$ and $\varphi_{(4)}$ are terms of the first, second, third, and fourth-order in the power series expansion of $\varphi$, $\mathbf{A}_{(1)}$ and $\mathbf{A}_{(2)}$ are terms of the first and second-order in the series expansion of $\mathbf{A}$. On computing (136), the following formula is taken into account:

$$\frac{e}{c} \cdot \frac{\partial b}{\partial t} \sim \frac{e^2}{m_0 c^3} b^2 \frac{\partial(\varphi^{ex} u^0)}{\partial t} \approx \frac{e^2}{m_0 c^3} b^2 \frac{\partial}{\partial t}\left[(\varphi^{ex}_{(1)} + \varphi^{ex}_{(2)} + \cdots)u^0\right] \tag{138}$$

where $\varphi^{ex}_{(1)}$ is the Coulomb potential which is a constant electric field, not dependent on the time, and so, $\partial \varphi^{ex}_{(1)}/\partial t = 0$ holds. As $\partial \varphi^{ex}_{(2)}/\partial t$ is the term of a high order of $1/c^3$, Eq. (138) is ignored. Hence, $b$ lies outside the differential symbol.

Next, we define the potentials of self-field giving rise to the radiation reaction as follows: if the interaction radius $R$ goes to zero, the potentials of the radiation reaction, $\varphi^{in}$ and $\mathbf{A}^{in}$, in (136) and (137) are defined as

$$\varphi^{in} = \lim_{R \to 0} \varphi \quad \text{and} \quad \mathbf{A}^{in} = \lim_{R \to 0} \mathbf{A}, \tag{139}$$

where $\varphi^{in}$ and $\mathbf{A}^{in}$ are the implicit functions expressed as $\varphi^{in} = \varphi^{in}(x, g)$ and $\mathbf{A}^{in} = \mathbf{A}^{in}(x, g)$.

Now, if one, according to the normalization rule 5 defined in Subsect. 6.2, normalizes $\varphi^{in}$ and $\mathbf{A}^{in}$, we have

$$\bar{\varphi}^{in} = \bar{\varphi}^{in}_{(1)} + \bar{\varphi}^{in}_{(2)} + \bar{\varphi}^{in}_{(3)} + \bar{\varphi}^{in}_{(4)} \tag{140}$$

and

$$\bar{\mathbf{A}}^{in} = \bar{\mathbf{A}}^{in}_{(1)} + \bar{\mathbf{A}}^{in}_{(2)}. \tag{141}$$

According to the normalization rule 1 (the normalization of field function, implicit function), we normalize $\varphi^{in}_{(1)}$ and $\mathbf{A}^{in}_{(1)}$. The normalized functions result in $\bar{\varphi}^{in}_{(1)} = \varphi^{in}_{(1)}(x, \bar{g})$ and $\bar{\mathbf{A}}^{in}_{(1)} = \mathbf{A}^{in}_{(1)}(x, \bar{g})$, respectively, with the normalized metric tensor $\bar{g}$; see (112), (113), (125), and (126). Hence,

$$\bar{A}^{in}_{\lambda(1)} = \lim_{R \to 0} \frac{1}{c} \cdot \frac{e \dot{x}_\lambda}{R\left[1 + 2\alpha \underline{A}^{in}_{\mu(1)} \underline{u}^\mu + 2\alpha \underline{A}^{ex}_{\mu} \underline{u}^\mu\right]} + C_\lambda \tag{142}$$

gives the first-order terms, $\varphi^{in}_{(1)}$ and $\mathbf{A}^{in}_{(1)}$:

$$\bar{\varphi}^{in}_{(1)} = \lim_{R \to 0} \frac{e}{R\left[1 + 2\alpha \left(\frac{e}{R} - \frac{e}{R} \cdot \frac{V^2}{c^2}\right) + 2\alpha \underline{A}^{ex}_{\lambda} \underline{u}^\lambda\right]} + C_0 = \frac{1}{2\alpha(1 - \beta^2)} + C_0 \tag{143}$$

and

$$\bar{\mathbf{A}}^{in}_{(1)} = \lim_{R \to 0} \frac{1}{c} \cdot \frac{e\mathbf{V}}{R\left[1 + 2\alpha \left(\frac{e}{R} - \frac{e}{R} \cdot \frac{V^2}{c^2}\right) + 2\alpha \underline{A}^{ex}_{\lambda} \underline{u}^\lambda\right]} + \mathbf{C}_1 = \frac{1}{c} \cdot \frac{\mathbf{V}}{2\alpha(1 - \beta^2)} + \mathbf{C}_1 \tag{144}$$

Substituting (143) and (144) in the action (133), we have:

$$\bar{A}^{in}_{\lambda(1)} \dot{x}^\lambda = \bar{\varphi}^{in}_{(1)} - \frac{1}{c} \bar{\mathbf{A}}^{in}_{(1)} \cdot \mathbf{V} = \frac{1}{2\alpha} + C_0 - \frac{1}{c} \mathbf{V} \cdot \mathbf{C}_1 \tag{145}$$

Now, if we find $C_0$ and $\mathbf{C}_1$ so that (145) leads to zero in the action (133), we have

$$\bar{A}^{in}_{\lambda(1)} \dot{x}^\lambda = 0. \tag{146}$$

with $\mathbf{C}_1 = \mathbf{0}$ and $C_0 = -1/2\alpha$. In the action (133) for the motion of a charge, the first-order term of the interaction of the charge with the self-field, i.e., $\bar{A}^{in}_{\lambda(1)} \dot{x}^\lambda$ vanishes. Thus, $\left(\bar{\varphi}^{in}_{(1)}, \bar{\mathbf{A}}^{in}_{(1)}\right)$ has no influence on the potential of radiation reaction. $\bar{\varphi}^{in}_{(2)}$ also vanishes because $e$ in $\bar{\varphi}^{in}_{(2)}$ is a constant and then the differentiation with respect to time leads to zero. Next, in $\bar{\varphi}^{in}_{(3)}$, according to the normalization rule 5 (corresponding to the series expansion), $A^{in}_{\lambda(1)}$ entered the denominator of $b$ should be replaced by already normalized potential $\bar{A}^{in}_{\lambda(1)}$ and then $\bar{A}^{in}_{\lambda(1)} \underline{u}^\lambda$ becomes zero according to (146). Accordingly, $\bar{\mathbf{E}}^{in}_{(3)}$ leads to

$$\bar{\mathbf{E}}^{in}_{(3)} = \lim_{R \to 0} \text{grad} \bar{\varphi}^{in}_{(3)} = b \frac{e}{2c^2} \ddot{\mathbf{n}} \bigg|_{R=0} \tag{147}$$

with

$$\dot{\mathbf{n}} = \frac{\partial}{\partial t} \frac{\mathbf{R}}{R}$$

where $b$ becomes 1 from $\frac{e}{c} \bar{A}^{in}_{\lambda(1)} \dot{x}^\lambda = 0$, and $\mathbf{n}$ is a unit vector of the radius $\mathbf{R}$ from the charge to a given point of the field. If $\mathbf{R}$ goes to zero, $\mathbf{n}|_{R=0}$ becomes a constant vector or zero, and then the differentiation of **n** with respect to time vanishes. Finally, $\bar{\varphi}_{(4)}^{in}$ does not go to zero. In the Maxwell theory, $\varphi_{(4)}$, by the gauge transformation, was transformed into zero; see p. 222 of Ref. 22. But in the action (133), as the gauge principle is not allowed, $\varphi_{(4)}$ cannot be transformed to zero. $b$ in $\bar{\varphi}_{(4)}^{in}$ is equal to 1 in the same way as in the case of $\bar{\varphi}_{(3)}^{in}$, and so, $\bar{\varphi}_{(4)}^{in}$ is the same as the fourth term of the power series in the Maxwell theory: the dependence of $\bar{\varphi}_{(4)}^{in}$ on the external field is ignored in our consideration because it includes the terms of higher order than $1/c^4$.

Now, we find $\mathbf{E}_{(4)}^{in}$ from $\bar{\varphi}_{(4)}^{in}$:

$$\mathbf{E}_{(4)}^{in} = -\mathrm{grad}\bar{\varphi}_{(4)}^{in} = \lim_{R \to 0} \frac{e}{6c^3} \cdot \frac{\partial^3}{\partial t^3}(\nabla R^2) = \lim_{R \to 0} \frac{e}{3c^3} \cdot \frac{\partial^3 \mathbf{R}}{\partial t^3} \tag{148}$$

Using

$$\mathbf{R} = \mathbf{R}_0 - \mathbf{r}, \quad \dot{\mathbf{R}} = -\dot{\mathbf{r}} = -\mathbf{V}, \quad \ddot{\mathbf{R}} = -\dot{\mathbf{V}}, \quad \dddot{\mathbf{R}} = -\ddot{\mathbf{V}},$$

(148) leads to

$$\mathbf{E}_{(4)}^{in} = -\mathrm{grad}\bar{\varphi}_{(4)}^{in} = -\frac{e}{3c^3}\ddot{\mathbf{V}}, \tag{149}$$

where $\mathbf{E}_{(4)}^{in}$ is not dependent on **R**, and so, the limit symbol in (149) vanishes.

Next, we find $\mathbf{E}_{(2)}^{in}$ from $\bar{\mathbf{A}}_{(2)}^{in}$:

$$\mathbf{E}_{(2)}^{in} = -\frac{1}{c}\frac{\partial \bar{\mathbf{A}}_{(2)}^{in}}{\partial t} = \frac{1}{c^3}e\ddot{\mathbf{V}}, \tag{150}$$

where $\mathbf{A}_{(2)}^{in}$ coincides with $\mathbf{A}_{(2)}$ given in (137), and the normalization of $\mathbf{A}_{(2)}^{in}$ was done in the same way as that of $\varphi_{(3)}^{in}$, with $b = 1$. Thus, the radiation reaction force $\mathbf{F}_R$ can be written as follows:

$$\mathbf{F}_R = \bar{e}\mathbf{E}_{(4)}^{in} + \bar{e}\mathbf{E}_{(2)}^{in} = -\frac{\bar{e}e}{3c^3}\ddot{\mathbf{V}} + \frac{\bar{e}e}{c^3}\ddot{\mathbf{V}} \approx \frac{2}{3c^3}\frac{e^2\ddot{\mathbf{V}}}{(1 + 2\alpha\underline{A}_\lambda^{ex}\underline{u}^\lambda)}, \tag{151}$$

The potentials of self-field contributed to the effective charge have been neglected in (151). Under some approximation, (151) coincides with that in the Maxwell theory. As shown in (151), the radiation reaction depends on the external field (with the nonlinear version) and contains the interference. The energy scale of a particle and the electromagnetic field is $b_0 = |2e\underline{A}_\lambda^{ex}\underline{u}^\lambda/m_0c^2|$. Under the approximation condition $b_0 < 1$, the expansion of the Taylor series in (151) arrives at

$$\mathbf{F}_R \approx \frac{2e^2\ddot{\mathbf{V}}}{3c^3}\left(1 - 2\frac{e}{m_0c^2}\underline{A}_\lambda^{ex}\underline{u}^\lambda\right) = \frac{2e^2\ddot{\mathbf{V}}}{3c^3} - \frac{4e^3\ddot{\mathbf{V}}}{3c^5}(\underline{A}_\lambda^{ex}\underline{u}^\lambda) \tag{152}$$

In (152), the second term gives interference effect, which is not in agreement with the LSP and gauge principle. (151) and (152) imply that while the interference effect can be enough neglected under the condition of weak field or low energy ($b_0 \ll 1$), under the strong field condition or the semi-strong field condition (when $b_0 \approx 1$ or $b_0$ is not small enough), the interference effect has significant influence on the motion of a charge. For example, consider an electron interacting with the strong electromagnetic field (e.g., high energy laser pulse). $\mathbf{F}_R$ is given by

$$\mathbf{F}_R \approx \frac{2}{3c^3} \frac{e^2 \ddot{\mathbf{V}}}{\left(1 + 2\frac{|e|}{m_0 c^3} \mathbf{A} \cdot \mathbf{V}\right)}, \tag{153}$$

where $\mathbf{A}$ is the vector potential of the plane wave and $\mathbf{V}$ is the velocity of the electron. In this case, $\mathbf{A}$ is the free electromagnetic field devoid of charges producing the field on the right side of the field equations. For such a free electromagnetic field, the Maxwell theory can be used as it is. It should be emphasized that the nonlinearity of the field equations is determined by the effective charge which depends on the self-field and external field and so yields the interference effect between the fields. Therefore, in the absence of the charge, the equations for free electromagnetic field are the same as in the Maxwell theory.

Supposing that $b_0 = |2e\mathbf{A}^l \cdot \mathbf{V}/m_0 c^3|$ is not small enough and $\mathbf{A} \cdot \mathbf{V} > 0$, (153) shows that the radiation reaction remarkably diminishes as compared to the case of the absence of interference effect, and thus, brings about energy loss smaller than the computation in the Maxwell theory. Recently, in QED, the topics of quantum recoil also, called quantum radiation reaction (QRR) attract a great deal of attention [24-37]. The phenomena such as particle beam spreading [38], cooling [39,40], and trapping [41,42] can be all elaborated in terms of quantum recoil experienced by particles interacting with laser pulses. The investigation of QRR confirms in a simulation study that while considering the radiation reaction on an electron interacting with a high energy laser pulse, the energy loss is significantly reduced in comparison with the classical prediction (the computation in the Maxwell theory) due to the quantum interference [43]: the correlation and interference between scattering events separated by arbitrarily large phase differences. Of course, quantum interference is a pure quantum mechanical phenomenon and is different from classical interference. But it is interesting to note that as argued above, the classical interference effect also yields such a reduction of energy loss as that arising from quantum interference. Our description of interference bears a drawback that does not take into account quantum effects. In future studies, when accounting for quantum effects, it is anticipated that the classical interference effect will give a significant impact on quantum theory. Finally, if one expresses the radiation reaction in the 4-covariant form, the equation of motion of a charge is given by

$$m_0 c \frac{du_\lambda}{ds} - \frac{1}{c} \bar{e} \frac{dA_\lambda}{ds} = \frac{1}{c} \bar{e}(F^{ex}_{\lambda\sigma} + F^{in}_{\lambda\sigma}) u^\sigma \tag{154}$$

with

$$F^{in}_{\lambda\sigma} \approx \begin{pmatrix} 0 & E^{in}_x & E^{in}_y & E^{in}_z \\ -E^{in}_x & 0 & 0 & 0 \\ -E^{in}_y & 0 & 0 & 0 \\ -E^{in}_z & 0 & 0 & 0 \end{pmatrix}$$

where $\bar{A}_\lambda = \bar{A}^{ex}_\lambda + \bar{A}^{in}_\lambda$, $E^{in}_i = -\partial \bar{\varphi}^{in}/\partial x_i - \partial \bar{A}^{in}_i/c\partial t$ and $\bar{A}^{in}_\lambda = (\bar{\mathbf{A}}^{in}, \bar{\varphi}^{in})$ given by $\bar{\varphi}^{in} \approx \bar{\varphi}^{in}_{(4)}$ and $\bar{\mathbf{A}}^{in} \approx \bar{\mathbf{A}}^{in}_{(2)}$. On the other hand, according to the normalization 3 ($\bar{\varepsilon} = \bar{\varepsilon}(x, \bar{F})$), (134) and (135) arrive at

$$\begin{cases} \bar{A}_\lambda = \frac{1}{c} \cdot \frac{e\dot{x}_\lambda}{R[1 + 2\alpha \bar{A}_\mu^{ex} \bar{u}^\mu]} \\ \bar{A}_\mu^{in} \bar{u}^\mu \approx 0 \end{cases} \quad (155)$$

where $\bar{A}_\mu^{ex} \approx \underline{A}_\mu^{ex}$. $\bar{u}^\mu$ is the normalized 4-velocity: for $\bar{u}^\mu$ is discussed in the next subsection.

In the case of a free particle, the action is given by

$$S = -m_0 c^2 \int dt \left(1 - \frac{V^2}{c^2}\right)^{\frac{1}{2}} \left(1 + 2\alpha \bar{A}_\lambda^{in} u^\lambda\right)^{\frac{1}{2}} = -m_0 c^2 \int dt \left(1 - \frac{V^2}{c^2}\right)^{\frac{1}{2}}, \quad (156)$$

where $\bar{A}_\lambda^{in} u^\lambda = 0$: it follows from the definition of the inertial reference system, according to which a *free particle cannot accelerate automatically to emit any radiation, and so the radiation reaction without the external field can never be present*. This conclusion is subject to the correspondence principle that the action for a free particle should coincide with that in the SR. In this context, it should be emphasized that the absurd result in the Maxwell theory according to which radiation damping without external field arises and so a charge automatically accelerates is no more valid.

In the description of Coulomb potential and radiation reaction we used the approximate field equation; see Sect. 6. Such a situation is not highly satisfactory. In fact, the left-hand side of the field equation is the same as in the Maxwell theory. Unfortunately, We did not find complete nonlinear field equations. But even though a complete nonlinear field equation is found, it has no essential influence on our description for the natural elimination of infinite quantities by using normalization. It is due to the fact that there exists a current of effective charges on the right-hand side of the field equation, and the functions of the potential fields are always expressed as implicit ones. The approach that eliminates divergences using the normalization rules cannot be altered but applied as it is. On the other hand, in the case of expanding potentials in a power series, with an approximation condition of $V/c \ll 1$, divergences are always manifested in only the first-order term of the expansion. This shows that even if a complete potential equation and solution is, in the future, founded, corrections newly added to the former solution will be reflected in high order terms that do not give rise to divergences. As a result, because divergences manifested in the first-order term are naturally removed by the normalization of implicit function, the current approach will be still valid for a complete nonlinear equation to be founded in the future.

### 8.3. Electrostatic energy of charges and total energy of particle-field

Here we show how the nonlinear character of interaction and the breaking of LSP and gauge symmetry bring about the finiteness of the total energy of particle-field. As the action is defined on KR space, the gauge symmetry breaking is inevitable; see Sect. 3 and Sect. 5. First of all, calculate the energy of the electrostatic field produced by charges. From (94), the energy-momentum tensor of the electrostatic field is given by

$$T_i^k = -\frac{1}{4\pi}\left(F_{i\lambda} F^{k\lambda} - \frac{1}{4} F_{lm} F^{lm} \delta_i^k\right). \quad (157)$$

The energy density of the electrostatic field is

$$T_0^0 = -\frac{1}{8\pi} E_i E^i, \tag{158}$$

where $i$ is the space components. With

$$E^i = -\frac{1}{\sqrt{-g}} E_i, \tag{159}$$

referring to (84), (158) leads to

$$T_0^0 = \frac{1}{8\pi} E_i E_i \frac{1}{\sqrt{-g}} \tag{160}$$

Hence, the energy of the electrostatic field arrives at

$$U = \int T_0^0 \sqrt{-g}\, dV = \frac{1}{8\pi} \int \frac{1}{\sqrt{-g}} (E_i)^2 \sqrt{-g}\, dV = -\frac{1}{8\pi} \int E_i \frac{\partial \varphi}{\partial x^i}\, dV$$

$$= -\frac{1}{8\pi} \int \partial_i (E_i \varphi)\, dV + \frac{1}{8\pi} \int dV \varphi \partial_i E_i. \tag{161}$$

According to the Gauss theorem, the first integral is equal to the integral of $E_i \varphi$ over the surface surrounding the integration volume. Since the first integral is taken over the whole space and the field is zero at infinity, it vanishes. For a system of point charges, Eq. (119) arrives at

$$\partial_i E_i = 4\pi \sum_a \bar{e}_a \delta(\mathbf{r} - \mathbf{r}_a) \tag{162}$$

Substituting Eq. (162) in the second integral of Eq. (161), the integral sum over the charges is given by

$$U = \frac{1}{2} \sum_a \bar{e}_a \varphi_a \tag{163}$$

where $\bar{e}_a$ is the effective charge and $\varphi_a$ is the potential of the field produced by all charges, at the point where the charge is located. Now, normalizing (163) according to the normalization rule 3 ($\bar{\varepsilon} = \bar{\varepsilon}(x, \bar{F})$: see (117), it follows

$$\bar{U} = \frac{1}{2} \sum_a \bar{\bar{e}}_a \bar{\varphi}_a. \tag{164}$$

Using (129), $\bar{\bar{e}}_a$ and $\bar{\varphi}_a$ are given by

$$\begin{cases} \bar{\bar{e}}_a = \dfrac{e_a}{\left[1 + 2\alpha \left(\bar{\varphi}_a{}^{ex}(r_a) + \bar{\varphi}_a{}^{in}(r_a)\right)\right]} \\ \bar{\varphi}_a{}^{in}(r_a) = 0 \end{cases} \tag{165}$$

with

$$\begin{cases} \bar{\varphi}_a = \sum_b \dfrac{e_b}{R_{ab}\left[1 + 2\alpha\left(\bar{\varphi}_b{}^{ex}(r_b) + \bar{\varphi}_b{}^{in}(r_b)\right)\right]} \\ \bar{\varphi}_b{}^{in}(r_b) = 0 \end{cases} \qquad (166)$$

where $e_b$ is the arbitrary charge in the system and $R_{ab}$ the distance between $e_a$ and $e_b$, $\bar{\varphi}_b{}^{in}(r_b)$ the self-field produced by the charge $e_b$, and $r_a$ and $r_b$ are the spatial coordinates which $e_a$ and $e_b$ are located respectively. Thus, on computing the energy of field produced by point charges in (164), the infinite terms arising from self-field vanish naturally, and only the terms of interaction energy dependent on a mutual distribution of particles remain.

Next, from (76), the energy density of the particles is given by

$$T_0^0 = \sum_a \frac{m_0^{(a)} c^2}{\sqrt{-g}} u_0^{(a)} u_{(a)}^0 \frac{ds}{dt} \delta(\mathbf{r} - \mathbf{r}_0), \qquad (167)$$

and

$$T = \int \sum_a \frac{m_0^{(a)} c^2}{\sqrt{-g}} u_0^{(a)} \delta(\mathbf{r} - \mathbf{r}_0) \sqrt{-g}\, dV = \sum_a m_0^{(a)} c^2 u_0^{(a)} \qquad (168)$$

The normalization of $u_0^{(a)}$, subject to the normalization rule 2, yields

$$\bar{u}_{(a)}^0 = \frac{\left[1 + 2\alpha(\bar{\varphi}_a^{in} + \bar{\varphi}_a^{ex})\underline{u}_{(a)}^0\right]^{1/2}}{\sqrt{1 - \beta_a^2}} \qquad (169)$$

where $\underline{u}_{(a)}^0$ is the time component of 4-velocity in Minkowskian space: $\underline{u}_{(a)}^0 = 1/\sqrt{1 - \beta_a^2}$. Allowing for $\bar{\varphi}^{in}\underline{u}^0 = 0$, (168) arrives at

$$\bar{T} = \sum_a m_0^{(a)} c^2 \bar{u}_{(a)}^0 = \sum_a \frac{m_0^{(a)} c^2 \left(1 + 2\alpha \bar{\varphi}_{(a)}^{ex} \underline{u}^0\right)^{1/2}}{\sqrt{1 - \beta_a^2}} \qquad (170)$$

Hence, we have the total energy of particles and fields:

$$E = \bar{T} + \bar{U} = \sum_a \frac{m_0^{(a)} c^2 \left(1 + 2\alpha \bar{\varphi}_{(a)}^{ex} \underline{u}^0\right)^{1/2}}{\sqrt{1 - \beta_a^2}} + \frac{1}{2} \sum_a \bar{\bar{e}}_a \bar{\varphi}_a \qquad (171)$$

Therefore, infinite terms arising from the self-field vanish naturally and only terms of interaction energy dependent on the mutual distribution of particles remain. From (171), the total energy of a free particle and its self-field is given by

$$\begin{cases} E = \bar{T} + \bar{U} = m_0 c^2 / \sqrt{1 - \beta^2} + U_0 = mc^2 \\ U_0 = 0 \end{cases} \qquad (172)$$

(172) is formally the same as the energy formula for a particle in SR, but entirely different in essential

content. In SR, $mc^2$ is the energy confined only to a particle but $mc^2$ in (172) is the energy which includes not only the energy of the particle but also the energy of the electrostatic field. This is the inevitable conclusion following from the starting postulate I: see Sect. 2. Thus, when we consider only the field energy, the energy of a static self-field (divergent in the Maxwell theory) naturally becomes zero. This is attributed to the fact that the energy of the static self-field is included in $mc^2$ according to the starting postulate I. However, (171) bears a drawback. We rewrite the formula for energy shown in (30)

$$E = \left(\frac{\bar{m}_0 c^2}{\sqrt{1-\beta^2}} - \bar{e}\varphi\right) + \bar{e}\varphi \tag{173}$$

In case of a system consisting of two charges (173) yields:

$$E = \left(\frac{\bar{m}_0^{(1)} c^2}{\sqrt{1-\beta_1^2}} - \bar{e}_1\varphi_2\right) + \frac{\bar{e}_1\bar{e}_2}{r} + \left(\frac{\bar{m}_0^{(2)} c^2}{\sqrt{1-\beta_2^2}} - \bar{e}_2\varphi_1\right) =$$

$$= \frac{\bar{m}_0^{(1)} c^2}{\sqrt{1-\beta_1^2}} + \frac{\bar{m}_0^{(2)} c^2}{\sqrt{1-\beta_2^2}} - \frac{\bar{e}_1\bar{e}_2}{r} \tag{174}$$

where $\varphi_1 = \bar{e}_1/r$, $\varphi_2 = \bar{e}_2/r$, and $\bar{e}_1\varphi_2 = \bar{e}_2\varphi_1 = \bar{e}_1\bar{e}_2/r$. Comparing (173) and (174) with (171), it is easily shown that (171) does not contain the term $-\bar{e}\varphi$ which is contributed to eliminating the additional energy that charges own jointly: see Sect.2. Such a difficulty is rooted in the fact that the action is the approximate version and so deriving the energy-momentum tensor, the term $\bar{e}dA_\lambda/cds$ giving birth to $(-\bar{e}\varphi)$ is not entirely taken accounted for. But, even though a complete action function for the field is found in the future, since the energy of self-field vanishes, it would hold intact that the total energy of a free particle and its self-field equals $mc^2$ and the total energy of a system consisting of charges and their fields also leads to a finite quantity with physical meaning. From (30), a complete expression for the total energy of the system becomes

$$E = \sum_a \left(\frac{m_0^{(a)} c^2 \left(1 + 2\alpha \bar{\varphi}_{(a)}^{ex} \underline{u}^0\right)^{1/2}}{\sqrt{1-\beta_a^2}} - \bar{\bar{e}}_a \bar{\varphi}_a\right) + \frac{1}{2}\sum_a \bar{\bar{e}}_a \bar{\varphi}_a \tag{175}$$

When taking the Newtonian approximation on (175), we can easily confirm that (171) is reduced to the formula for energy in Newton mechanics: for reference, see (33) and (34). Now, if one obtains the total energy of the system under the semi-strong field condition, the drawback mentioned above does not arise. Under the semi-strong field condition, the derivation of the conservation law of energy-momentum proceeds from the action function (43) and the action function (92) for the field. The detailed derivation is the same as in the Maxwell theory: see Sect.33 of Ref. 22. The normalization procedure is also the same as in (164)-(166). The result is

$$E = \sum_a \frac{m_0^{(a)} c^2}{\sqrt{1-\beta_a^2}} + \frac{1}{2}\sum_a \underline{\bar{e}}_a \bar{\varphi}_a \tag{176}$$

where $\underline{e}_a$ denotes the semi-effective charge and $\underline{\bar{e}}_a$ the normalized semi-effective charge. (176) also provides the conclusion that the energy of the self-field naturally vanishes and the total energy of the system necessarily leads to the finite. Under relevant approximation, the normalized semi-effective charge $\underline{\bar{e}}_a$ is reduced to a constant charge $e$ and so (176) leads to the formula for energy in the Maxwell theory.

The above argument implies the following.

First, LSP and gauge symmetry are basic factors giving birth to divergences in the Maxwell theory.

Second, CNE built on KR space in which accompanies the breakings of LSP and gauge symmetry helps us to obtain finite physical quantities: the finite total energy of particles and their fields, and the elimination of divergent terms in radiation reaction.

## 9. Quantum nonlinear electrodynamics

Classical theory underlies quantum theory and hence as long as the basis of classical theory varies, quantum theory should also be, from a new viewpoint, rethought and rebuilt naturally. In this section, we briefly consider quantum nonlinear electrodynamics (QNE) based on classical nonlinear electrodynamics.

### 9.1. Modification of Dirac equation

First of all, rewrite the formula (42) for the total energy of a particle in CNE, as discussed in Sect. 3:

$$E = [\bar{m}_0 c^4 + c^2(P)^2]^{1/2} \quad (177)$$

where $\bar{m}_0$ is the effective inertial mass: see (28). Comparing (177) with the formula for energy in the Maxwell theory, in (177) appear the effective inertial mass, instead of a constant mass in the Maxwell theory. Namely, if one replaces the constant mass in the Maxwell theory with the effective inertial mass, the formula for the total energy of a particle formulated on KR space is obtained.

Now, for a reformulation of quantum electrodynamics, we suppose that this relationship holds as it is. Consequently, by replacing mass in the Dirac equation by effective inertial mass, the modified version of the Dirac equation is obtained:

$$[\gamma^\mu i \partial_\mu - \bar{m}_0] \psi(x) = 0 \quad (178)$$

When obtaining Eq. (178), the detailed derivation is omitted because it is too obvious. As easily understood, Eq. (178), due to the effective inertial mass $\bar{m}_0$, constitutes a nonlinear equation: $\bar{m}_0 = m_0 (1 + 2\alpha A_\mu u^\mu)^{\frac{1}{2}}$. Hence, finding an analytical solution to Eq. (178) is a very difficult problem. In this context, discuss an approximate version of Eq. (178). For this purpose, we employ (48):

$$E = \left[m_0^2 c^4 + c^2 \left(\mathbf{P} - \frac{e}{c}\mathbf{A}\right)^2\right]^{\frac{1}{2}} + \underline{e}\varphi, \quad (179)$$

with

$$\underline{e} = \frac{e}{(1 + 3\alpha A_\mu \underline{u}^\mu)}.$$

From (179), it follows that

$$[\gamma^\mu(i\partial_\mu - \underline{e}A_\mu) - m_0]\psi(x) \approx 0. \tag{180}$$

That is, if one replaces the charge, a constant in the Dirac equation by the semi-effective charge, Eq. (180) is obtained. Owing to the approximate version Eq. (180), the modified Dirac equation is divided into two parts, i.e., the free particle term and the interaction term, and therefore, all approaches of QED (linear QED) can be used as it is. The interaction Hamiltonian $H_s$ and the Lagrangian $L$ are given by

$$H_s = \int L dx^3 \tag{181}$$

and

$$L = \underline{e}\bar{\psi}\gamma_\lambda A^\lambda \psi. \tag{182}$$

To express $\underline{e}$ as a quantum mechanical operator, transform $3eA_\lambda u^\lambda/m_0c^2$ in the denominator of $\underline{e}$ as follows:

$$\frac{3}{m_0c^2}eA_\lambda \underline{u}^\lambda = \frac{3}{m_0c^2}\underline{e}\left(1 + \frac{3}{m_0c^2}eA_\lambda \underline{u}^\lambda\right)A_\lambda \underline{u}^\lambda = \alpha_0 \underline{e}A_\lambda \dot{x}^\lambda \tag{183}$$

with

$$\alpha_0 = \frac{\frac{3}{m_0c^2}\left(1 + \frac{3}{m_0c^2}eA_\lambda \underline{u}^\lambda\right)}{\sqrt{1-\beta^2}} \approx \frac{\frac{3}{m_0c^2}\left(1 + \frac{2}{m_0c^2}eA_\lambda \underline{u}^\lambda\right)^{3/2}}{\sqrt{1-\beta^2}} \approx \frac{3}{m_0c^2}\frac{(g_{\mu\nu}\delta^{\mu\nu}/4)^{3/2}}{\sqrt{1-\beta^2}}, \tag{184}$$

where $g_{\mu\nu}\delta^{\mu\nu} = 4(1 + \frac{2}{m_0c^2}eA_\lambda \underline{u}^\lambda)$ and $\left(1 + \frac{2}{m_0c^2}eA_\lambda \underline{u}^\lambda\right)^{3/2} = (g_{\mu\nu}\delta^{\mu\nu}/4)^{3/2} \approx 1 + \frac{3}{m_0c^2}eA_\lambda \underline{u}^\lambda$ was used: see (98). If one expresses (183) as a quantum mechanical operator, the result is

$$\frac{3}{m_0c^2}eA_\lambda \underline{u}^\lambda = \alpha_0 \underline{e}A_\lambda \dot{x}^\lambda \rightarrow \int dx^3 \alpha_0 \underline{e}\bar{\psi}\gamma_\lambda A^\lambda \psi. \tag{185}$$

Thus, (182) arrives at

$$L = \underline{e}\bar{\psi}\gamma_\lambda A^\lambda \psi = \frac{e\bar{\psi}\gamma_\lambda A^\lambda(x,g)\psi}{1 + \int dx^3 \alpha_0 \underline{e}\bar{\psi}\gamma_\lambda A^\lambda(x,g)\psi} = \frac{L_m}{1 + \int dx^3 \alpha_0 L} \tag{186}$$

where

$$\underline{e} = \frac{e}{1 + \int dx^3 \alpha_0 \underline{e}\bar{\psi}\gamma_\lambda A^\lambda \psi} \tag{187}$$

and $L_m = e\bar{\psi}\gamma_\lambda A^\lambda \psi$. Therefore, $L$ becomes an implicit function, and so do $H_s$ in (181). We have

$$H_s = \int dx^3 \frac{L_m}{1 + \int dx'^3 \alpha_0 L} = \int dx^3 \frac{L_m}{1 + H'_s} \tag{188}$$

The implicit function $H_s$ becomes the key to solving divergence problems.

### 9.2. Normalization of S-matrix and its convergence

In this subsection, we define a modified version of the S-matrix in quantum nonlinear electrodynamics, by using the normalization rules, normalize the S-matrix, and then discuss its convergence. For the S-matrix treated in QED, the $n$th-order term of the perturbation expansion is as follows:

$$S^{(n)} = \frac{(-i)^n}{n!} \int dx_1^4 \cdots dx_n^4 T\big(H_s(x_1) \cdots H_s(x_n)\big), \tag{189}$$

In this case, as $H_s$ is given by (188) derived from the modified Dirac equation (180), (189) is different from that in QED.

At present, in order to normalize $S^{(n)}$, express (189) as a concise form of implicit function. For this, first of all, we consider $H'_s = \int d\,'x^3 \alpha_0 L$ which is present in the denominator of $\underline{e}$ in (188). According to the mean value theorem in mathematics, there exists $t_0$ satisfying

$$\int dt\, H'_s = T_0 H'_s|_{t=t_0} = T_0 \bar{H}'_s, \tag{190}$$

where $\bar{H}'_s$ is the average value of $H'_s$, $T_0$ the time interval between $t_1$ (before the interaction occurs) and $t_2$ (after the interaction disappears), and $T_0$ is supposed to be finite.

Now suppose that the following approximation condition holds:

$$\bar{H}'_s = H'_s|_{t=t_0} \approx H'_s. \tag{191}$$

The reason is as follows. In the scattering theory, it is generally supposed that an interaction time is very short (for example, the annihilation of electron-positron, the electromagnetic interactions in the vicinity of a nucleus, etc.). In this case, (191) is valid, and thus, (190) and (191) yield

$$H'_s \approx \bar{H}'_s = \frac{1}{T_0} \int dx^4\, \alpha_0 L(x). \tag{192}$$

Now, if we substitute (192) in (188), (188) arrives at

$$H_s = \int dx^3 \frac{L_m}{1 + \frac{1}{T_0} \int dx'^4 \alpha_0 L(x)} \tag{193}$$

Substituting (193) in (189), (189) leads to

$$S^{(n)} = \frac{(-i)^n}{n!} \bar{e}_0^n \int dx_1^4 \cdots dx_n^4 T\big(L'_m(x_1) \cdots L'_m(x_n)\big) \tag{194}$$

where $L'_m = \bar{\psi}\gamma_\lambda A^\lambda \psi$ and

$$\bar{e}_0^n = \frac{e^n}{\left[1 + V_{(n-1)} + (\alpha_0)^n \frac{1}{(T_0)^n} S^{(n)}\right]}.$$

In the denominator of $\bar{e}_0^n$, $S^{(n)}$ is the $n$th-order $S$-matrix and $V_{(n-1)}$ is a sum of all possible products consisting of the respective combination taken by an arbitrary number (from 1 to $n-1$) out of a set of $\frac{1}{T_0}(-i)\alpha_0 \int dx_j^4 L(x_j)$ from $j=1$ to $j=n-1$: for example, in the case of $n=4$, $V_{(3)}$ can be written by

$$V_{(3)} = \frac{(-i)^3}{3!}\left(\frac{\alpha_0}{T_0}\right)^3 \int dx_1^4 dx_2^4 \, dx_3^4 T\big(L(x_1)L(x_2)L(x_3)\big) + \frac{(-i)^2}{2!}\left(\frac{\alpha_0}{T_0}\right)^2 \int dx_1^4 dx_2^4 \, T\big(L(x_1)L(x_2)\big) +$$

$$+ \frac{(-i)^2}{2!}\left(\frac{\alpha_0}{T_0}\right)^2 \int dx_2^4 dx_3^4 \, T\big(L(x_2)L(x_3)\big) + \frac{(-i)^2}{2!}\left(\frac{\alpha_0}{T_0}\right)^2 \int dx_1^4 dx_3^4 \, T\big(L(x_1)L(x_3)\big) +$$

$$+(-i)\left(\frac{\alpha_0}{T_0}\right)\int dx_1^4\, L(x_1) + (-i)\left(\frac{\alpha_0}{T_0}\right)\int dx_2^4\, L(x_2) + (-i)\left(\frac{\alpha_0}{T_0}\right)\int dx_3^4\, L(x_3).$$

where every term consists of the respective combination taken by an arbitrary number (from 1 to $n-1$). $(\alpha_0)^n$ can be enough considered as a constant: when normalizing $S^{(n)}$, $\big(g_{\mu\nu}\delta^{\mu\nu}/4\big)^{3n/2}$ entered $(\alpha_0)^n$ becomes 1 and from (192), $\beta^2$ can be viewed as a constant within a very short time of the interaction. Therefore, when one normalizes the $S$-matrix, from the beginning, these are put in front of $S^{(n)}$ to avoid troublesome calculations. (194) can be written in a more concise form:

$$S^{(n)} = \frac{(-i)^n}{n!} \cdot \frac{S_0^{(n)}}{1+(\alpha_0)^n \frac{1}{(T_0)^n} S^{(n)} + V_{(n-1)}}, \qquad (195)$$

where $S_0^{(n)} = e^n \int dx_1^4 \cdots dx_n^4 T\big(\bar{\psi}\gamma_\lambda A^\lambda \psi \cdots \bar{\psi}\gamma_\lambda A^\lambda \psi\big)$ is the same as that in QED.

In QED, the renormalization procedure is used for eliminating the divergent terms appearing in the S-matrix. As well known, the simplest method by which separates finite and infinite parts in the integral formula is to expand a Taylor series on the external momentum. For example, the Taylor series expansion of $\Gamma(P^2)$ in a neighborhood of $P^2 = 0$ is given by

$$\Gamma(P^2) = a_0 + a_1 P^2 + \cdots + \frac{1}{n!}a_n (P^2)^n + \cdots \qquad (196)$$

with the coefficients

$$a_n = \frac{\partial^n}{\partial P^2}\Gamma(P^2)|_{P^2=0},$$

where the coefficients $a_n$ with $n \geq 1$ are finite, and only $a_0$ diverges logarithmically. If one expresses a sum of all finite quantities as $\tilde{\Gamma}(S)$, (196) leads to

$$\Gamma(S) = \Gamma(0) + \tilde{\Gamma}(S), \qquad (197)$$

where $\Gamma(0)$ is an infinite quantity and $\tilde{\Gamma}(S)$ a finite quantity. This situation is similar to the case where a divergent term occurs in the Maxwell theory. Actually in the Maxwell theory, in case of expanding the potential of the field as a series in powers of $R/c$ and considering radiation damping, the first term $e/R$ of the series expansion diverges and the second term $\mathbf{A}'^{(2)}$ in the series expansion of vector potential expressed by partial differentiation becomes finite; see Subsect. 8.2. Such a situation allows the Maxwell theory to subtract artificially the infinite quantity $\left(\lim_{R\to 0} e/R\right)$, regarding it as a meaningless thing, or under an excuse according to which within a small spatial area of $R \approx 10^{-13}$cm, not classical physics but quantum theory is essential to "solve" such theoretical difficulty. Then renormalization procedure was also introduced to solve this difficulty [6,15]. The situation in which divergences in QED occur within a small area (a region of large momentum) and the first term in the Taylor series expansion leads to an infinite quantity, which is formally the same as the difficulty in the Maxwell theory. This implies that th divergence occurring in quantum electrodynamics is rooted in the difficulty of classical electrodynamics.

Now in the paper, see how brief and concise the divergence problem of the S-matrix is solved without the renormalization which is so complicated and artificial. In the first place, expand the Taylor series of $S^{(n)}$ and divide it into divergent and measurable finite terms. In this case, by expanding the Taylor series of $S_0^{(n)}$ included in $S^{(n)}$, we can obtain the sum of the divergent term $S_{in}^{(n)}$ and the finite term $S_f^{(n)}$:

$$S^{(n)} = S_{in}^{(n)} + S_f^{(n)}. \tag{198}$$

The divergent term $S_{in}^{(n)}$ and the finite term $S_f^{(n)}$ are respectively given by

$$S_{in}^{(n)} = \frac{S_{0(in)}^{(n)}}{1 + (\alpha_0)^n \frac{1}{(T_0)^n}\left(S_{in}^{(n)} + S_f^{(n)}\right) + V_{(n-1)}} + C \tag{199}$$

and

$$S_f^{(n)} = \frac{S_{0(f)}^{(n)}}{1 + (\alpha_0)^n \frac{1}{(T_0)^n}\left(S_{in}^{(n)} + S_f^{(n)}\right) + V_{(n-1)}} + C, \tag{200}$$

where $C$ is a finite constant, $S_{0(in)}^{(n)}$ the sum of all divergent terms in the $n$th-order term of the scattering matrix expansion in QED, and $S_{0(f)}^{(n)}$ the convergent term (a finite correction).

Next, by using the normalization rules discussed in Subsect. 7.2, we obtain a measurable finite correction from $S^{(n)}$. According to the normalization rule 4, the normalization of a quantity represented as a series expansion becomes a sum of the normalizations of each physical quantity that constitute it. From this, it follows that

$$\bar{S}^{(n)} = \bar{S}_{in}^{(n)} + \bar{S}_f^{(n)}. \tag{201}$$

We first find $\bar{S}_{in}^{(n)}$. If one, in terms of the normalization rule 1, transforms $g_{\mu\nu}$ (the metric tensor of KR space) into $\delta^{\mu\nu}$ (the Minkowski metric tensor) in the implicit function $S_{in}^{(n)}$, $S_{in}^{(n)}$ and $S_f^{(n)}$ in the denominator of (199) leads to $S_{0(in)}^{(n)}$ and $S_{0(f)}^{(n)}$ defined in QED. And $V_{(n-1)}$ is replaced by the already normalized quantity $\bar{V}_{(n-1)}$. The scattering matrix is normalized in serial order starting from the lowest order terms, and therefore, $\bar{V}_{(n-1)}$ should be considered as the already normalized term: see the normalization rule 5. As a result, we have the normalized quantities

$$\begin{cases} g_{\mu\nu} \to \delta_{\mu\nu} \\ \alpha_0 \underline{e}\bar{\psi}\gamma_\lambda A^\lambda \psi \approx \dfrac{3}{m_0 c^2}\underline{e}\bar{\psi}\gamma_\lambda A^\lambda \psi \to \dfrac{3}{m_0 c^2} e\bar{\psi}\gamma_\lambda \, \underline{A}^\lambda \psi \\ A^\lambda(x,g) \to \underline{A}^\lambda(x); \text{field function in the Maxwell theory} \\ S_{in}^{(n)} \to S_{0(in)}^{(n)} \\ S_f^{(n)} \to S_{0(f)}^{(n)} \\ V_{(n-1)} \to \bar{V}_{(n-1)} \end{cases} \quad (202)$$

Hence, (199) arrives at

$$\bar{S}_{in}^{(n)} = \dfrac{S_{0(in)}^{(n)}}{1+\left(\dfrac{\alpha_0}{T_0}\right)^n \left(S_{0(in)}^{(n)} + S_{0(f)}^{(n)}\right) + \bar{V}_{(n-1)}} + C \quad (203)$$

If one divides the numerator and denominator by $S_{0(in)}^{(n)}$ and takes into account the fact that $S_{0(in)}^{(n)}$ is an infinite quantity, (203) leads to

$$\bar{S}_{in}^{(n)} = \left(\dfrac{\alpha_0}{T_0}\right)^{-n} + C \quad (204)$$

By choosing properly the integral constant $C$, (204) must be

$$\bar{S}_{in}^{(n)} = 0 \quad (205)$$

Next, we obtain $\bar{S}_f$. According to the first and fifth normalization rules, the following transformation is done for $S_f^{(n)}$ in (200):

$$g_{\mu\nu} \to \delta_{\mu\nu}, \quad S_f^{(n)} \to S_{0(f)}^{(n)}, \quad S_{in}^{(n)} \to \bar{S}_{in}^{(n)} = 0, \quad V_{(n-1)} \to \bar{V}_{(n-1)}. \quad (206)$$

Therefore, (200) leads to

$$\bar{S}_f^{(n)} = \dfrac{S_{0(f)}^{(n)}}{1+\left(\dfrac{\alpha_0}{T_0}\right)^n S_{0(f)}^{(n)} + \bar{V}_{(n-1)}} \approx S_{0(f)}^{(n)}. \quad (207)$$

Taking together (205) and (207), the result is

$$\bar{S}^{(n)} = \bar{S}_{in}^{(n)} + \bar{S}_f^{(n)} = \bar{S}_f^{(n)} \approx \bar{S}_{0(f)}^{(n)}. \quad (208)$$

The method discussed above can be applied to all terms of the $n$th order formula in the scattering matrix expansion, and then as an inevitable consequence, one can obtain a measurable finite quantity with a physical meaning. Here, the detailed calculations for individual terms of the S-matrix are omitted because the result is very obvious.

Now, by applying the approach mentioned above to a detailed example, consider how a divergence is naturally solved according to a logical procedure. When one considers one-loop correction in which the quantized field $A^\mu$ is inserted as the first order, $\underline{e}\gamma_\mu A^\mu$ is transformed as follows:

$$\underline{e}\gamma_\mu A^\mu \to \underline{e}A^\mu\big(\gamma_\mu + \Lambda_\mu + i\Pi_{\mu\nu}iD^{\nu\iota}\gamma_\iota\big), \tag{209}$$

where $\Lambda_\mu$ is the vertex function and $\Pi_{\mu\nu}$ the vacuum polarization, $D^{\nu\iota}$ the propagation function of a free photon. Now, if one expands $\Lambda_\mu$ as a Taylor series in $\hat{P} = m$, the result is

$$\Lambda_\mu = \Lambda_{\mu(\hat{P}=m)} + \Lambda_{\mu(f)} = \Lambda_{in} + \Lambda_f, \tag{210}$$

where $\Lambda_{\mu(\hat{P}=m)}$ is a logarithmically divergent term and $\Lambda_{\mu(f)}$ a finite term. On the other hand, the series expansion for $\Pi_{\mu\nu}$ in $k^2 = 0$ leads to

$$\Pi_{\mu\nu} = \left(\delta_{\mu\nu} - \frac{k_\mu k_\nu}{k^2}\right)k^2 \Pi_{(k)} \tag{211}$$

with

$$\Pi_{(k)} = \Pi_{(0)} + \Pi_{(k^2)(f)} = \Pi_{in} + \Pi_f,$$

where $\Pi_{(0)}$ is a logarithmically divergent term and $\Pi_{(k^2)(f)}$ a finite term. Therefore, the Hamiltonian of the interaction of an electron with the external field is given by

$$H = \underline{e}\int d^3 r \bar{\psi}_{(x)}\left[\gamma_\mu + (\Lambda_{in} + \Lambda_f) + i\left(\delta_{\mu\nu} - \frac{k_\mu k_\nu}{k^2}\right)k^2(\Pi_{in} + \Pi_f)iD^{\nu\iota}\gamma_\iota\right]\psi(x) = H_{in} + H_f, \tag{212}$$

where $H_{in}$ is a divergent term and $H_f$ a finite term. And the semi-effective charge is

$$\underline{e} = \frac{e}{1 + \alpha_0 \int d^3 r' \bar{\psi}_{(x)}\underline{e}A^\mu(\gamma_\mu + \Lambda_\mu + i\Pi_{\mu\nu}iD^{\nu\iota}\gamma_\iota)\psi_{(x)}} = \frac{e}{1 + \alpha_0(H_{in} + H_f)} \tag{213}$$

Thus, the Hamiltonian arrives at

$$H = H_{in} + H_f = \frac{H'_{in}}{1 + \alpha_0(H_{in} + H_f)} + \frac{H'_f}{1 + \alpha_0(H_{in} + H_f)} \tag{214}$$

where $H'$ is the term that includes the constant charge $e$. Now, if one normalizes (214) according to the first and fourth normalization rules, it follows that

$$\bar{H} = \bar{H}_{in} + \bar{H}_f \tag{215}$$

Referring to (201) - (208), if one calculates (215), the following result is easily obtained:

$$\bar{H} = \frac{H_{0(f)}}{1 + \alpha_0 H_{0(f)}} \approx H_{0(f)}(1 - \alpha_0 H_{0(f)}) \approx H_{0(f)}, \tag{216}$$

where $H_{0(f)}$ is the interaction Hamiltonian of the electron with the external field which has a finite value, obtained in QED.

As shown above, in QNE, without doing such artificial and complicated calculations as renormalization, divergences are, easily and naturally, removed and finite quantities are obtained. If so, where does this key comes from? The key, in a word, consists in the introductions of feedback of interaction between particle and field, and of implicit function, mathematical expression of feedback.

As already mentioned above, an implicit function $F$ has the following structure:

$$F \sim \frac{F_0}{1 + \alpha_0 F}, \tag{217}$$

where $F_0$ is a physical quantity defined in Minkowski space. As the scattering matrix and the Hamiltonian for interaction are the implicit functions, a divergence of the function $F_0$, in the way included as a variable owing to a feedback, induces another divergence of the function $F$ (in the denominator of (217)). Then, since the two divergences are combined as the relationship of numerator and denominator, the divergences are naturally eliminated with normalization. Consequently, the introductions of feedback and implicit function are very essential to eliminate divergences which are knotty points in QED.

Summarizing all the arguments mentioned above arrives at the following conclusions.

First, from the scattering matrix of nonlinear quantum electrodynamics, we obtained the finite corrections with the procedure in which infinite quantities are removed naturally and by themselves. From this, we can make correct theoretical analyses of the variation of Coulomb law, Lamb shift, and anomalous magnetic moment of an electron.

Second, the above consideration was made on the basis of CNE established on KR space. The experimental verifications of the theoretical results mentioned above confirm the validity of CNE in which nonlinear quantum electrodynamics is rooted.

## 9.3. Electron scattering around a nucleus

In Maxwell's electrodynamics, the field produced by a charge does not depend on fields created by any other charges, and the fields produced by every charge are independent of each other without any intervention. From this, the principle of linear superposition is established and the gauge symmetry holds. But in the paper, the field produced by a charge depends on the external field, and then the principle of linear superposition and gauge symmetry are destroyed; see Sect. 5. Consequently, the electromagnetic interaction comes to have a nonlinear character. To see this, we consider an electron scattering in the external electrostatic field in view of QNE. As shown in (129), the electrostatic field produced by a nucleus is given by

$$\varphi_p = \frac{Ze}{r\left[1 + 2\frac{Ze}{Mc^2}\bar{\varphi}_e\right]} \approx \frac{Ze}{r\left[1 + 2\frac{Ze}{Mc^2}\underline{\varphi}_e\right]} \tag{218}$$

where $Z$ is the number of protons and $M$ the mass of the nucleus, $\underline{\varphi}_e$ the potential produced by the electron in the Maxwell theory, $\bar{\varphi}_e$ the normalized potential: $\bar{\varphi}_e \approx \underline{\varphi}_e$.

The energy of interaction between an electron and a nucleus is

$$U = \bar{e}\varphi_p = \frac{e}{\left[1 + 2\frac{e}{mc^2}\underline{\varphi}_p u^0\right]} \cdot \frac{Ze}{r\left[1 + 2\frac{e}{Mc^2}\underline{\varphi}_e\right]} \approx \frac{Ze^2}{r\left[1 + 2\frac{e}{mc^2}\underline{\varphi}_p\right]} = \frac{Ze^2}{r\left[1 - Z\frac{r_0}{r}\right]} \tag{219}$$

using

$$\left|\frac{e}{mc^2}\varphi_p\right| \gg \left|\frac{Ze}{Mc^2}\varphi_e\right|$$

where $m$ is the mass of the electron, $r_0$ the electron radius, and $\varphi_p$ the potential produced by the nucleus in the Maxwell theory.

Now, from (219), consider a new correction relevant to an additional nonlinear effect obtained by the term $Z\frac{r_0}{r}$. The formula for scattering cross-section well known in QED is as follows [16]:

$$d\sigma = \frac{tr[(\hat{p}+m)\gamma^0(\hat{p}'+m)\gamma^0]}{2}\left|e\tilde{\varphi}_p(q)\right|^2\delta(E'-E)dE'd\Omega, \qquad (220)$$

where $E$ is the energy of an electron, $q = p - p'$ the variation in the momentum of the electron produced by the scattering, and $\tilde{\varphi}_p(q)$ the Fourier component of the external field. Then we obtain $\tilde{\varphi}_p(q)$ using the Fourier transformation

$$\tilde{\varphi}_p(q) = \int d^3r\, e^{iq\left(1-\frac{\bar{q}}{q_0}\right)r}\frac{Ze}{r-Zr_0} = \frac{Ze}{q^2\left(1-\frac{\bar{q}}{q_0}\right)^2}e^{iq\left(1-\frac{\bar{q}}{q_0}\right)Zr_0}, \qquad (221)$$

leading to

$$\left|e\tilde{\varphi}_p(q)\right|^2 = \frac{Z^2e^4}{q^4\left(1-\frac{\bar{q}}{q_0}\right)^4}, \qquad (222)$$

where $q_0$ is a maximum variation in the momentum of the electron produced by the scattering, and $\bar{q}$ is an average variation. The average variation in electron momentum is expressed as

$$\bar{q} = \int d^3r\, \bar{\psi}q\psi \qquad (223)$$

In (222), $q$ and $q_0$ can be represented as

$$q^2 = 4|p^2|\sin^2(\theta/2) = 4E^2v^2\sin^2(\theta/2) \text{ and } q_0 = 2Ev\sin(\pi/4) = \sqrt{2}Ev \qquad (224)$$

Then, we have

$$\frac{\bar{q}}{q_0} = \frac{\bar{q}}{\sqrt{2}Ev}\ ,$$

where $v$ is the velocity of an incident electron and $\theta$ the scattering angle. When the scattering angle $\theta$ arrives at $\pi/2$, the average variation of electron momentum $\bar{q}$ reaches the maximum value $q_0$. In this consideration, the condition $\theta > \pi/2$ is not valid. The condition $\theta = \pi/2$ and $\bar{q} = q_0$ become singular points in which electron capture around a nucleus occurs (Figure 1).

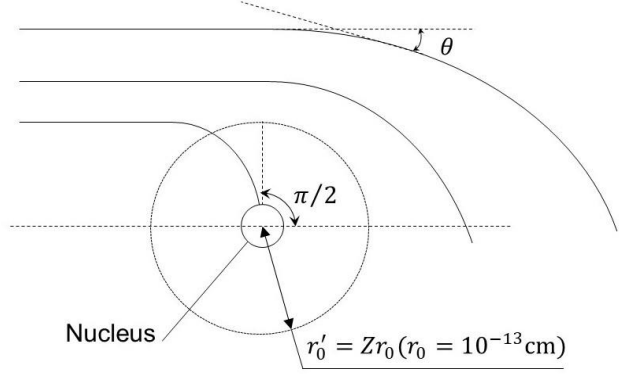

Figure 1. The scattering and capture of electrons by a nucleus

Substituting (224) in (222), we have

$$\left|e\tilde{\varphi}_p(q)\right|^2 = \frac{Z^2 e^4}{16E^4 v^4 \sin^4(\theta/2)}\left(1 - \frac{\bar{q}}{\sqrt{2}Ev}\right)^{-4} \tag{225}$$

Then, substituting (225) in (220), the result is

$$\frac{d\sigma}{d\Omega} = \frac{Z^2 e^4\left(1 - v^2\sin^2(\theta/2)\right)}{4E^4 v^4 \sin^4(\theta/2)}\left(1 - \frac{\bar{q}}{\sqrt{2}Ev}\right)^{-4} \tag{226}$$

In QED, a formula for the scattering cross-section is

$$\frac{d\sigma}{d\Omega} = \frac{Z^2 e^4\left(1 - v^2\sin^2(\theta/2)\right)}{4E^4 v^4 \sin^4(\theta/2)} \tag{227}$$

Comparison of (227) with (226) shows that the term $\left(1 - \frac{\bar{q}}{\sqrt{2}Ev}\right)^{-4}$ appears newly. This new term implies a nonlinear effect of interaction. When a beam of electrons passes through a thin target, an average distance of the interaction between a nucleus and electrons in the beam can be defined. Then $\bar{q}$ is related to the average distance of interaction as follows:

$$-d\bar{q}_{(\bar{r})} = \alpha_s \bar{q}_{(\bar{r})} d\bar{r}, \tag{228}$$

where $\bar{r}$ is the average distance of interaction and $\alpha_s$ $\left(a_s = \frac{1}{Zr_0}\right)$ is a constant defined from the approximate condition, $\bar{r} \gg Zr_0$. (228) yields

$$\bar{q}_{(\bar{r})} = q_0 e^{-\alpha_s \bar{r}\left(1 - \frac{Zr_0}{\bar{r}}\right)} \tag{229}$$

Substitution of (229) into (226) arrives at

$$\frac{d\sigma}{d\Omega} = \frac{Z^2 e^4\left(1 - v^2\sin^2(\theta/2)\right)}{4E^4 v^4 \sin^4(\theta/2)}\left(1 - e^{-\alpha_s \bar{r}\left(1 - \frac{Zr_0}{\bar{r}}\right)}\right)^{-4} \tag{230}$$

(230) shows that for experiments allowing $\bar{r} \gg Zr_0$, the term $e^{-\alpha_s \bar{r}\left(1-\frac{Zr_0}{\bar{r}}\right)}$ can be enough neglected, and thus, (230) leads to (227). However, when a major part of the interactions occurs in the neighborhood of $Zr_0$, that is, in the case of $\bar{r} \approx Zr_0$, $e^{-\alpha_s \bar{r}\left(1-\frac{Zr_0}{\bar{r}}\right)}$ cannot be neglected. Supposing that an electron arrives at the average distance of interaction with a nucleus, $\bar{r} = Zr_0$, the term $\left(1 - e^{-\alpha_s \bar{r}\left(1-\frac{Zr_0}{\bar{r}}\right)}\right)^{-4}$ becomes zero and (230) goes to infinity. In order for the scattering cross-section to have a finite value, the charge $e$ in the numerator of (230) must vanish: see (59). That is, in a singular point of $\bar{r} = Zr_0$, the electron is captured by the nucleus, and then vanishes. But in the former QED, the scattering cross-section is independent of the average distance of interaction, and any possibility of an electron capture around a nucleus is not derived. The experimental verification for (230) will prove that the nonlinear effect of electromagnetic interaction actually arises, and the principle of linear superposition and gauge symmetry are not valid. Until now we have discussed several main things of classical and quantum nonlinear electrodynamics.

## 10. Conclusion and outlook

We considered classical and quantum nonlinear electrodynamics without renormalization. As shown in Sect. 2, from the starting postulate I and the correspondence principle, we could get the final conclusion that CNE must be necessarily studied; see Sect. 2. The action in Maxwell's theory includes only external fields due to the divergence of the self-field potential, and with the aid of the approach of renormalization, the interaction of a particle with its self-field is, additionally from the outside of a logical system of the theory, introduced to the equations of motion of the particle.

We formulated the action functions for the motion of a particle and fields and found the equation system for that, which includes the interactions of the particle with not only the external fields but also the self-field from the beginning, in a more general framework. From the equation system, we presented a theoretical analysis of the mechanism of the annihilation and production of particles, based on the energy conservation, and demonstrated that the results agree well with the experiments under the strong field condition $|U/m_0 c^2| \cong 1$ ($U$ ; interaction energy), which is clearly absent in Maxwell's theory. We confirmed that CNE brings about a more universal and inclusive conservation law of charge, manifested in special forms according to the different energy scales of interactions. It was also shown that the gauge symmetry has an application limit, which holds under the weak field condition (in Maxwell's theory). Obtained based upon the starting postulate I and correspondence principle, the action for the motion of a charge and field was inevitably formulated on KR space with implicit function metric and so all functions characterizing particle and field also were expressed as implicit functions. We showed that such implicit function was a mathematical manifestation of feedback coming from the interaction (self-interaction) of a charge and its self-field. By introducing the normalization rules (starting postulate II) formulating the physical meaning of implicit function, divergence problems arising in self-field energy and radiation reaction were naturally and easily solved. This implies that divergences in Maxwell's theory do not arise due to "pointness" of a particle but the limitations of an approximate version of a theory that holds only within some limit of application (week field or low-energy). In the motion equation of a charge, charge and inertial mass, regarded as the constants in Maxwell's theory, were replaced by effective charge and effective inertial mass dependent on the scalar product of 4-field and 4-velocity.

A basic signature of nonlinear theory different from linear theory consists in a manifestation of the interference effect with which the field produced by a particle is dependent on an external field and then the interaction between particles interferes with each other. In CNE, from the dependence of the effective charge and effective inertial mass on the 4-field and 4-velocity, an interference effect necessarily arises. From the action formulated on KR space, we found the modified versions of Coulomb potential and radiation action, followed by the interference effects. We demonstrated that while considering radiation reaction for an electron interacting with the strong electromagnetic field, the interference brought about radiation reaction energy loss smaller than the computation in Maxwell's theory, which will give a significant impact on quantum theory. A fatal drawback of Maxwell's theory consists in the divergence of the electrostatic field energy of charges. It was found that the electrostatic field energy of charges obtained from the energy-momentum tensor is finite and the total energy of a particle and its self-field is the same as $mc^2$, subject to the starting postulate I.

Finally, based upon CNE, we briefly considered quantum nonlinear electrodynamics in which the modified version of the Dirac equation was derived and the divergence problem coming from the S-matrix was naturally solved without renormalization. Throughout this paper, we demonstrated that only classical and quantum electrodynamics can yield finite physical quantities without invoking renormalization and with sensible mathematics.

The obtained results can be extended at least in three directions. First, the nonlinear equation of the electromagnetic field should be studied in a more complete form, satisfying an accurate conservation formula for the total energy-momentum of particle and field. Second, nonlinear quantum effects arising at regimes of semi-high energy or high energy, including quantum interference waiting to prove experimentally, should be further discovered. Third, the formulation of CNE on KR space including Riemann space in a special form enables us to study the unification of electromagnetism and gravitation. In this context, the unification should be considered: in this case, according to the correspondence principle, a unified theory should draw all theoretical results suggested by Maxwell's theory and GR, and should present new theoretical results relevant to the mutual dependence between electromagnetism and gravitation to be experimentally verified in the future.